\documentclass[conference,compsoc]{IEEEtran}
\usepackage[utf8]{inputenc}
\usepackage{amsmath,amssymb,amsfonts}
\usepackage{graphicx}
\usepackage[tight,footnotesize]{subfigure}
\usepackage{multirow}
\usepackage{fancyvrb}
\usepackage[flushleft, online]{threeparttable}
\usepackage{xurl}
\usepackage{array}
\usepackage{algorithmic}
\usepackage{booktabs}
\usepackage{xcolor}
\usepackage{makecell}
\usepackage{pifont}
\usepackage{ulem}
\usepackage[
    colorlinks,
    linkcolor=blue,
    anchorcolor=blue,
    citecolor=blue,
    urlcolor=blue
]{hyperref}
\usepackage{textcomp}

\usepackage{cite}

\newcommand{\subject}[1]{\vspace{3pt}\noindent\textbf{#1}}
\newcommand{\subsubject}[1]{\vspace{3pt}\noindent\textit{#1}}

\newcommand\malurl[1]{\href{notalink}{{\nolinkurl{#1}}}}
\newcounter{finding}
\newcommand{\finding}[1]{
\vspace{3pt}
\noindent
\framebox{
\begin{minipage}[b]{3.3in}
\noindent \textbf{Finding \Roman{finding}}: \textit{#1}
\stepcounter{finding}
\end{minipage}
}
\vspace{-5pt}
}
\renewcommand{\finding}[1]{}

\newcommand{\ignore}[1]{}

\newcolumntype{L}[1]{>{\raggedright\let\newline\\\arraybackslash\hspace{0pt}}m{#1}}
\newcolumntype{C}[1]{>{\centering\let\newline\\\arraybackslash\hspace{0pt}}m{#1}}
\newcolumntype{R}[1]{>{\raggedleft\let\newline\\\arraybackslash\hspace{0pt}}m{#1}}

\begin{document}
\title{Dissecting Open Edge Computing Platforms: Ecosystem, Usage, and Security Risks}

\IEEEoverridecommandlockouts
\author{\IEEEauthorblockN{
Yu Bi\IEEEauthorrefmark{1}\IEEEauthorrefmark{2}\thanks{\IEEEauthorrefmark{2}Both authors contributed equally to this research.}, 
Mingshuo Yang\IEEEauthorrefmark{3}\IEEEauthorrefmark{2}, 
Yong Fang\IEEEauthorrefmark{1}, 
Xianghang Mi\IEEEauthorrefmark{1}\IEEEauthorrefmark{4}\thanks{\IEEEauthorrefmark{4}Corresponding authors.},
Shanqing Guo\IEEEauthorrefmark{3}\IEEEauthorrefmark{5}\IEEEauthorrefmark{4}, 
Shujun Tang\IEEEauthorrefmark{6} 
and Haixin Duan\IEEEauthorrefmark{6}\IEEEauthorrefmark{7}
}

\IEEEauthorblockA{
\IEEEauthorrefmark{1}University of Science and Technology of China, 
\IEEEauthorrefmark{3}Shandong University,
}
\IEEEauthorblockA{
\IEEEauthorrefmark{5}Shandong Key Laboratory of  Artificial Intelligence Security,
}
\IEEEauthorblockA{
\IEEEauthorrefmark{6}Qi An Xin Technology Research Institute, 
\IEEEauthorrefmark{7}Tsinghua University
}
\IEEEauthorblockN{\small
\IEEEauthorrefmark{1}\{biu, yongf\}@mail.ustc.edu.cn, xmi@ustc.edu.cn,\\
\IEEEauthorrefmark{3}yangmingshuo@mail.sdu.edu.cn, guoshanqing@sdu.edu.cn,\\
\IEEEauthorrefmark{6}tangshujun@qianxin.com,
\IEEEauthorrefmark{7}duanhx@tsinghua.edu.cn
}
}

\maketitle

\begin{abstract}
  Emerging in recent years, open edge computing platforms (OECPs) claim large-scale edge nodes, the extensive usage and adoption, as well as the openness to any third parties to join as edge nodes. For instance, OneThingCloud, a major OECP operated in China, advertises 5 million edge nodes, 70TB bandwidth, and 1,500PB storage. However, little information is publicly available for such OECPs with regards to their technical mechanisms and involvement in edge computing activities. Furthermore, different from known edge computing paradigms, OECPs feature an open ecosystem wherein any third party can participate as edge nodes and earn revenue for the contribution of computing and bandwidth resources, which, however, can introduce byzantine or even malicious edge nodes and thus break the traditional threat model for edge computing. In this study, we conduct the first empirical study on two representative OECPs, which is made possible through the deployment of edge nodes across locations, the efficient and semi-automatic analysis of edge traffic as well as the carefully designed security experiments. As the results, a set of novel findings and insights have been distilled with regards to their technical mechanisms, the landscape of edge nodes, the usage and adoption, and the practical security/privacy risks. Particularly, millions of daily active edge nodes have been observed, which feature a wide distribution in the network space and the extensive adoption in content delivery towards end users of 16 popular Internet services. Also, multiple practical and concerning security risks have been identified along with acknowledgements received from relevant parties, e.g., the exposure of long-term and cross-edge-node credentials, the co-location with malicious activities of diverse categories, the failures of TLS certificate verification, the extensive information leakage against end users, etc. 
\end{abstract}

\section{Introduction}
\label{sec:intro}
The paradigm of edge computing has been proposed and experimented for many years~\cite{cicirelli2017edge, taleb2017mobile, motlagh2017uav, liu2019edge, alonso2020intelligent, zhang2021emp}, while realistic and operational edge computing platforms tend to be small-scaled~\cite{xu2021cloud}. In recent years, multiple \textit{open} edge computing platforms (OECPs) emerge and get increasing adoption, which feature large-scale and widely-distributed edge nodes, real-world and extensive adoption, and the openness to any third parties to join as edge nodes and earn revenue for the contribution of computing and bandwidth resources. For instance, 
OneThingCloud (also branded as Xingyu Cloud)~\cite{xycloud,onethingcloud}, one of the two representative OECPs under our study, claims to have 5 million edge nodes, 70Tbs bandwidth capacity, and 1,500PB storage capacity, as well as the adoption by many popular online services, e.g., Kuaishou, iQIYI, Bilibili.  

However, these OECPs are  black boxes since little is publicly available regarding their technical mechanisms or real-world usage. Furthermore, it is intuitive to wonder what practical security/privacy implications these OECPs may render, especially considering the claimed extensive adoption by online services.  For instance, our preliminary investigation reveals that none of these OECPs enforce strict vetting for edge node operators, and any third party can turn their devices into edge nodes, which can introduce byzantine or even malicious edge nodes.  It also breaks the traditional threat model for edge computing  wherein edge nodes are fully controlled by the ECP and are usually considered as trusted. 
% Also, such in-production ECPs enable the co-location of edge computing tasks on the same edge node, and co-located edge computing tasks can belong to different parties and may interfere with each other when the isolation is not well enforced. 
To clear away the dense mist surrounding OECPs, we conduct an extensive security study on two representative ones, namely, TipTime~\cite{tiptime} and OneThingCloud~\cite{onethingcloud}.
As elaborated in \S\ref{sec:background}, these two stand out in a comparative analysis of all identified OECPs. 
Our study would not be possible without the design and implementation of a set of automatic or semi-automatic tools. Particularly, to capture edge activities and edge traffic, a deployment framework is designed to deploy edge nodes across OECPs at different locations and capture their edge traffic. 
Then, given edge traffic of multiple TBs, an efficient edge traffic analyzer is built up to concurrently process edge traffic flows, group them into pre-defined traffic categories, as well as correlating them with various real-world parties. Also, a security testbed is carefully designed to empirically evaluate security risks of OECPs while avoiding interfering with real-world edge computing activities. Applying these tools leads to new understandings of OECPs with regards to their technical mechanisms, the landscape of edge nodes, real-world usage, as well as security and privacy risks. 

Our key findings are highlighted as below. % \bullet 
First of all, Both OECPs share many properties in their technical mechanism and service model. Briefly, both  enable the co-deployment of different edge computing tasks on a single edge node and each edge computing task runs like an independent kingdom inside the edge node. Also, when it comes to edge node recruitment, both support a variety of computing devices including mobile phones, desktop/laptop computers, computing servers, WIFI hotspots and routers, or even network attached storage devices. And the support of so many heterogeneous device types has likely contributed to the large scale of edge nodes. Also, the ecosystem of OECPs involve a complicated interactions among three types of participants:  edge node operators, OECP providers, and tenants of edge computing tasks. Among these participants, edge node operators transform their devices into edge nodes and contribute computing/bandwidth resources for revenue. Then, ECP providers are responsible for coordinating edge nodes and distributing edge computing tasks across edge nodes, while edge computing tenants subscribe to OECPs so as to deploy edge computing tasks.  

Also, edge nodes of both OECPs turn out to be large-scaled and widely distributed. Leveraging the edge node deployment framework, we have captured over 6TB edge traffic, among which, counterpart edge nodes attached to 22,214 IP addresses are observed to have communicated with ones under our control. Although being far lower-bound estimated for the scale of edge nodes, these edge nodes present a wide distribution in 12,935 different /24 IPv4 network blocks and 67 autonomous systems. Furthermore, due to the involvement in content delivery as CDN servers, edge nodes of both ECPs tend to be resolved to fully qualified domain names (FQDNs) that match general patterns. Querying passive DNS datasets with these FQDN patterns reveals that 34,364,400 IP addresses have ever served as edge nodes of either OECPs between the time period of January 2021 and November 2023. And the two OECPs have a total of millions of daily active edge nodes observed. Furthermore, regarding the usage and adoption of 
OECPs, both OECPs are designed to be agnostic to edge computing tasks and thus generalize to any edge computing tasks. However, as observed from the edge traffic,  they are currently mainly used for content delivery tasks which involve 6 different CDNs and 16 popular content providers, e.g.,  Douyin, KuaiShou, Bilibili, Netease, TouTiao news, etc. Also, the content payloads delivered by edge nodes are of diverse categories, which include not only traditional static web files (Javascript and media files), but also emerging content payloads, e.g., program files and machine learning models.

Furthermore, we have located a set of practical security/privacy risks and most of them can be attributed to the flawed designs of either the two OECPs or the CDN modules running inside the edge nodes. First of all, many edge node IPs suffer from low threat reputation, i.e., they have been involved in various malicious activities.  For instance, as revealed by a proprietary and ever-updating threat intelligence platform operated by a major security vendor, over 68\% edge node IPs have been involved in 10 or more malicious traffic flows during the time period between January 2022 and November 2023. Also, almost 4\% have been reported by VirusTotal as either hosting malicious URLs or distributing various malware payloads. The low reputation of edge node IPs can likely disrupt legitimate on-edge activities such as delivering content towards end users. 

Another security risk is the exposure of long-term and cross-edge credentials to potential attackers. As edge nodes hosting content delivery tasks can serve as TLS servers, we find that edge nodes across platforms tend to share and locally store the same set of long-term TLS credentials (TLS private keys), which renders a non-negligible surface for MITM attacks against TLS traffic. Particularly, as long as one of these edge nodes was by-default malicious or compromised, TLS credentials for all these edge nodes would be leaked to the attackers, and transparent MITM attacks can be conducted. On the other hand, when serving as TLS clients (e.g., pulling control instructions from an edge server), edge nodes across the two OECPs fail to validate the server-side TLS certificates for part if not all of the TLS traffic flows. The negative impact of such validation failures can be well magnified by the large scale of edge nodes as well as the critical roles of these traffic flows in control-plane edge-to-server communication. Other security/privacy issues we have located include the insufficient confinement for co-located edge computing tasks, the extensive leakage of end-users' information (e.g., the device type, IP addresses and plaintext content), and the recruitment of edge nodes without verifiable authorization from device owners, etc. We have responsibly disclosed these security findings to relevant parties including OECP operators and CDN services. By this writing, OneThingCloud and Xingyu CDN have fully acknowledged our reports and are working to fix these issues. Besides, we have also provided several recommendations to address these security risks, e.g., the Tor-like edge-based routing protocol to mitigate the privacy risks against end users.  For more details, please refer to \S\ref{sec:security_disclosure}.

\subject{Ethical considerations.}
We take ethics seriously and have carefully designed our methodology to avoid any ethical issues.
Firstly, while collecting and measuring edge nodes and edge traffic, we conduct only statistical analysis to reveal scale and activity categories, avoiding any attempts to look into the content of plaintext traffic. Then, when evaluating identified security risks, we carefully design the experiments to avoid disrupting any content delivery activities. Also, all the identified security risks have been responsibly reported to the related parties, along with full acknowledgement received from OneThingCloud (\S\ref{sec:security_disclosure}).
On the other hand, we sought an IRB review before conducting this study, but unfortunately found out that the IRB at our institution has a limited scope of reviewing biological/medical studies and has yet to be capable of reviewing applications from other domains, including cybersecurity. 
Instead, we sought ethics reviews from multiple external cybersecurity researchers and have amended our methodology design by following their feedbacks.

Our contributions can be briefly summarized as below. Firstly, we conduct the first of its kind extensive study on two representative open edge computing platforms (OECPs). Then, our study has distilled a set of previously unknown findings on OECPs with regards to their technical mechanisms, landscape of edge nodes, usage and adoption, and practical security and privacy risks. Lastly, our study has provided a set of insights and recommendations for making future OECP designs more robust and secure. 
% The ecosystem of bulk sms services
% with a focus on their service model, functionalities, price policies, reseller programs
\section{Preliminaries}
\label{sec:background}
\subject{Open edge computing platforms (OECPs).}
Manually searching the web reveals a set of OECPs listed in Table~\ref{tab:comprison_with_OECP}. For each of them, we dive deeper to learn its service model, deployment options, scale, etc. Finally, we choose OneThingCloud and TipTime considering multiple factors. On one hand, OneThingCloud advertises largest volume of edge nodes. Although TipTime has not publicly disclosed its number of nodes on its website, the TipTime App~\cite{tiptimeDeployment} indicates that 127,000 device owners have joined the platform to run edge nodes, which is a remarkably large volume. On the other hand, they are the only two OECPs that support deployment through Docker, which eases our edge node deployment. One thing to note is that we have also identified OECPs from countries other than China, including Acruast~\cite{acurast} and Render Network~\cite{rendernetwork}. However, both of these OECPs have not disclosed their edge-node volume, along with certain deployment limitations. For instance, Acruast only supports mobile phones as nodes, while Render Network focuses solely on leveraging idle GPU resources.

\begin{table}
    \footnotesize
    \centering
        \caption{A comparison of OneThingCloud and TipTime with other OECPs.}
    \label{tab:comprison_with_OECP}
    \scalebox{0.75}{
    \begin{threeparttable}
    \begin{tabular}{cccccc}
        \toprule
    OECP Provider& Nodes& Storage& Bandwith& Customers\tnote{1}&Docker\tnote{2}\\
        \midrule
OneThingCloud~\cite{onethingcloud} & 5000K~\cite{onethingcloudNodesCustomers} & 1500PB~\cite{onethingcloudNodesCustomers}  & 70T~\cite{onethingcloudNodesCustomers} & 22~\cite{onethingcloudNodesCustomers} & ~\ding{52}\\
TipTime~\cite{tiptime} & N/A\tnote{3} & N/A & N/A & 6~\cite{tiptime} & ~\ding{52}\\
PPIO~\cite{PPIO} & 110K~\cite{PPIO} & N/A & 30T~\cite{PPIO} & 13~\cite{PPIO} & ~\ding{56} \\
JingXiang Cloud~\cite{JingXiangCloud} & 10K~\cite{JingXiangCloud}& 200PB~\cite{JingXiangCloud} & 100T~\cite{JingXiangCloud} & 1~\cite{JingXiangCloud} & ~\ding{56}\\
DianXin Cloud~\cite{DianXinCloud} & 3000K~\cite{DianXinCloud} & 200PB~\cite{DianXinCloud} & 100T~\cite{DianXinCloud} & 6~\cite{DianXinCloudCustomers} & ~\ding{56}  \\
XingSong Cloud~\cite{XingSongCloud} & 2000K~\cite{XingSongCloud} & 200PB~\cite{XingSongCloud} & 20T~\cite{XingSongCloud} & N/A & ~\ding{56} \\
Kun TAKE Cloud~\cite{KUNTAKECloud} & 100K~\cite{KUNTAKECloud} & 100PB~\cite{KUNTAKECloud} & 10T~\cite{KUNTAKECloud} & 6~\cite{KUNTAKECloud} & ~\ding{56} \\
Acurast~\cite{acurast} & N/A & N/A & N/A & N/A & ~\ding{56} \\
Render Network~\cite{rendernetwork}  & N/A & N/A & N/A & N/A & ~\ding{56} \\
         \bottomrule
    \end{tabular}
            \begin{tablenotes}
            \item [1] Customers denotes enterprises collaborating with OECPs and deploying edge tasks.
            \item [2] Docker denotes whether OECP allows nodes to be deployed using Docker.
             \item [3] N/A denotes the OECP didn't provide the data.
        \end{tablenotes}
    \end{threeparttable}}
\end{table}

\subject{Passive DNS and IP Intelligence}. 
    In our study, passive DNS plays an important role, especially when capturing IP addresses of edge nodes and measuring their temporal evolution. The passive DNS dataset adopted in our study comes from an industry collaborator which deploys sensors on DNS resolvers that are widely distributed in China. And its DNS sensors can observe a daily average of 600 billion DNS queries/responses, with the resulting passive DNS dataset covering 800 million unique domain names and 70 million unique IP addresses. Also, to profile the distribution of edge node IPs, we query a well-adopted IP intelligence service, namely, IPinfo~\cite{ipinfo}, which allows us to learn for each IP, its geolocation, country, city, autonomous system number (ASN), ISP, etc.
\section{Collecting and Analyzing Edge Activities}
\label{sec:method}
To fulfill our research goals of understanding OECPs and studying their potential security and privacy risks, the first step is to collect and analyze edge-node activities. This is enabled by two effective tools: an edge node deployment framework designed to run edge nodes and capture edge traffic, and an edge traffic analyzer designed to identify edge traffic signatures and distinguish edge traffic of different categories. Below, we provide more details for both tools. 

% EC node runner
\subsection{The Edge Node Deployment Framework}
\label{subsec:method_node_deployment}
% purpose
In a nutshell, this framework allows us to automatically deploy nodes across OECPs, capture their activities (e.g., network traffic), and 
backup the resulting measurement data.% 

Both TipTime and OneThingCloud have offered multiple options for deploying their edge nodes, and the full list of deployment optionscan be found in Appendix~\ref{appendix:deployment}.
In our deployment framework, we choose the option of Docker images since it allows quick and scalable deployment across heterogeneous devices.
% resource allocation to trigger traffic 
To support incoming connections that are either TCP or UDP,  we configure each edge container with the \textit{host} network type, which allows it to have full access to the network of the host,  a virtual machine deployed in a public cloud with static public IP addresses attached. 
Also, as illustrated in official tutorials~\cite{tt_revenue, wxy_revenue},  there is a minimum disk storage requirement which varies across OECPs. Specifically, OneThingCloud requires a minimum disk storage of 50GB in 2022 and 200GB in 2023, while it is 32GB for edge nodes of TipTime.
% Why and how to configure the network, so as to trigger effective EC traffic, especially usage traffic
Thus, when first deploying edge nodes in 2022, we configured each edge node with 100GB disk storage, 1 virtual CPT cores, 1GB memory, and 20Mbps downstream/upstream bandwidth. Then, as OneThingCloud upgraded the minimum disk storage requirement from 50GB to 200GB in 2023, we updated the configuration as well.
Also, tcpdump is instructed to capture its raw network traffic. 

% deployment
\subject{Deployment.} 
Considering both OECPs are dedicated to the China market, we deployed all the edge nodes in China. Also, to avoid geographical biases and enable a direct comparison among edge nodes in different locations, our edge nodes were deployed in three of the largest cities in China, namely, Beijing, Shanghai, and Shenzhen, which are also widely distributed in the geographical space. Also, our deployment consists of two phases. The first phase is between October 11, 2022, and December 10, 2022, which spanned 61 days involving 6 distinct edge node instances. Then, in June 2023, we noticed that some edge computing tasks may have got missed due to the constrained resources of our edge computing nodes. In another word, these edge computing tasks have a resource requirement higher than that of our initial edge computing nodes, e.g., a higher bandwidth of 30Mbps rather than 20Mbps is required along with a disk storage of 400GB rather than 100GB.  We thus enhanced the deployment configuration and deployed edge nodes for both OECPs for one more week in the location of Beijing around July 2023. 
Besides,  both OECPs provide a dashboard for the edge node operator to view and manage edge nodes under its control. Leveraging such a dashboard, we were allowed to verify that all deployed edge nodes had generated valid edge computing activities.
    
In summary, our deployment has involved 378 node-days (equivalent to running a single edge node for 378 days),  which has generated 57,547,905 traffic flows and 6990.94 GB network traffic. Also, during the deployment, our edge nodes have communicated with 1,210,761 distinct IP addresses, which are widely distributed in 1,016 autonomous systems and 105 countries. More details of this traffic dataset will be presented in \S\ref{sec:ecosystem} along with an in-depth analysis of its underlying activities which turn out to be content delivery activities involving a diverse set of CDN services and content providers.

\subsection{The Edge Traffic Analyzer}
\label{subsec:method_traffic_analyzer}
% - Categorize traffic flows by the functionalities a traffic flow serves

% - The  participants (organizations and Internet users) in edge computing activities 

\begin{figure}
    \centering    
    \includegraphics[width=1.0\columnwidth]{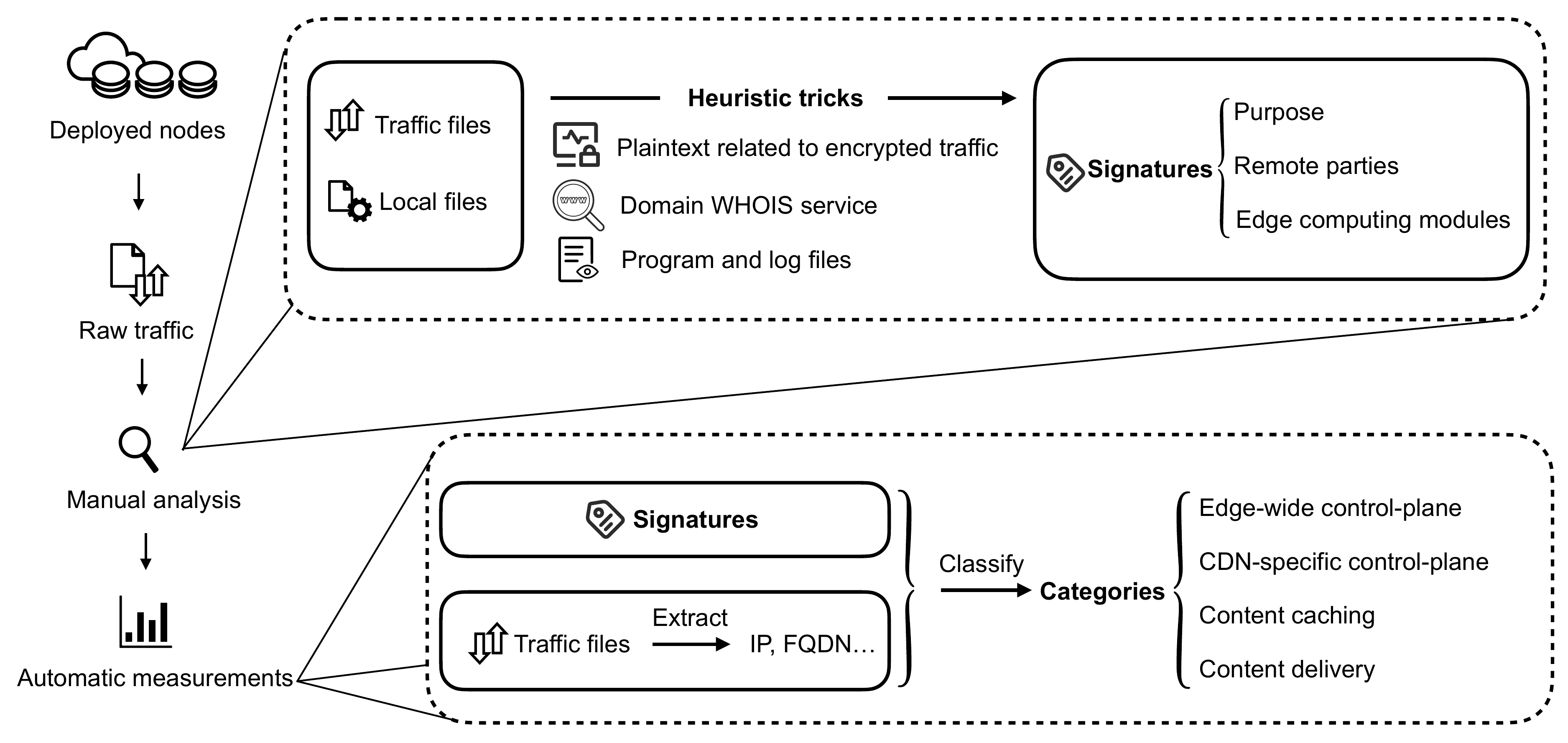}
    \caption{The pipeline of the edge traffic analyzer.
    }
    \label{fig:edge_traffic_analyzer}
\end{figure}

Given the large volume of edge traffic, we would like to figure out what purpose each traffic flow is intended for, reveal what remote parties have communicated with our edge nodes, and ultimately understand what edge computing activities have been conducted and through what kinds of network protocols. We pursue these tasks through a combination of manual analysis and automatic measurements. As shown in Figure~\ref{fig:edge_traffic_analyzer}, the manual analysis allows us to gain qualitative knowledge such as the categories of edge traffic flows, and the signatures to associate traffic flows with different categories or distinct remote parties. On the other hand, automatic measurements are designed to generate quantitative measurement results, e.g.,  the volume and shares of different traffic categories.

\subject{Manual analysis.} During our manual analysis, Wireshark was utilized to view the raw network traffic. However, it turns out to be non-trivial to locate the purposes of each traffic flow or the real-world remote parties underpinning them, especially when a traffic flow is encrypted. 
Fortunately, we have identified several effective tricks to conquer this challenge, including utilizing the correlation among traffic flows especially the correlation between plaintext traffic flows and encrypted ones, mapping domain names to their real-world entities by querying credible domain WHOIS dataset, and observating the difference between downstream and upstream traffic. More details can be found in Appendix~\ref{appendix:analyze}.

Leveraging a combination of these tricks and techniques, we have gained with high confidence a good understanding of both OECPs with regards to their technical mechanisms (i.e., network protocols), edge computing tasks, and the signatures to group traffic flows. Particularly, all edge computing tasks observed in our edge nodes turn out to be content delivery tasks that involve various CDN services and content providers. Also, a typical edge node tend to concurrently host multiple content delivery tasks that belong to different parties. 
More details will be presented in \S\ref{sec:ecosystem}.

\subject{Automatic measurements}. Given the qualitative understandings for the edge traffic along with a set of traffic signatures, an automatic measurement pipeline is built up to efficiently analyze all the captured edge traffic. This pipeline takes PCAP files of edge traffic as the input, and concurrently parses and analyzes traffic flows. For each traffic flow, important attributes (e.g., IP address, payload size, domains and URLs if any) will be extracted and traffic signatures will be matched against so as to group the traffic flow into pre-defined traffic categories as well as correlating it to real-world parties (e.g., specific CDN services, and content platforms).Below, we provide more details about this automatic analysis pipeline. 

When ingesting the raw PCAP files, the dpkt library\footnote{https://dpkt.readthedocs.io/en/latest} is utilized to facilitate fast packet parsing. Then, when extracting the remote addresses of a given traffic flow, 
the IP address and the transport-layer port can be easily extracted, which, however cannot reveal much human-readable information (e.g., what activities the flow is used for). Instead, the fully qualified domain name (FQDN) and the URL to which a traffic flow is intended to visit can be more helpful in terms of understanding its activities. 
For HTTP traffic, both the FQDN and the URL can be easily extracted from the plaintext payload. Furthermore, for TLS connections, we first try to extract the domain name from the server name indicator (SNI) extension of TLS. For other non-TLS but encrypted connections, we first build up mappings between domains and IPs, through parsing DNS responses recorded in the same PCAP file. Then, a TCP flow that is initiated by our edge node will be considered to have the domain name identified only when its remote IP address can be uniquely mapped to a domain name as recorded in DNS responses.   

Then, when it comes to assigning a traffic flow to different categories, aforementioned manual analysis has identified signatures to group traffic into the following 4 main categories: edge-wide control-plane, CDN-specific control-plane, content caching, and content delivery. Among these categories, edge-wide control-plane traffic encompasses the traffic flows between edge nodes and control servers of the respective OECP, which are aimed to fulfill goals including pulling control instructions, pushing logging data, and downloading executable payloads of edge computing tasks, etc. Then, as all edge computing tasks that we have observed are content delivery tasks, the second traffic category is CDN-specific control-plane traffic between an on-edge CDN module and the respective remote CDN control server. Then, across content delivery tasks, two more categories of traffic flows are observed to fulfill content delivery. One is the flows to cache content from either upstream content servers or counterpart CDN nodes, which we name as \textit{the content caching traffic}. The other category involves traffic flows that are intended to delivery content payloads to either end devices (e.g., viewers of videos) or counterpart CDN nodes, and we thus call it \textit{the content delivery traffic}. Then, since multiple CDN modules can be co-deployed on the same edge node, traffic flows of the same  CDN-relevant category can be further correlated with a specific CDN service (e.g., YunFan), or even a specific content provider (e.g., Douyin).
% Regarding the pre-defined categories, they are edge-wide control plane traffic, edge-wide data plane traffic, CDN-specific traffic for caching content from upstream content servers, CDN-specific traffic for delivering content to end users. Then, 
In a nutshell, the traffic signatures identified through manual  analysis enable us to group each traffic flow into these four main categories as well as correlating most content delivery/caching traffic flows with respective CDN service and content provider. For instance, we consider a content delivery traffic flow belongs to YunFan CDN and is used to delivery content to end users of the Douyin platform as long as its FQDN matches the pattern of \textit{*yf*.free-lbv6.idouyinvod.com}. 
\section{The Ecosystem}
\label{sec:ecosystem}

\begin{figure}
    \centering
    \includegraphics[width=.9\columnwidth]{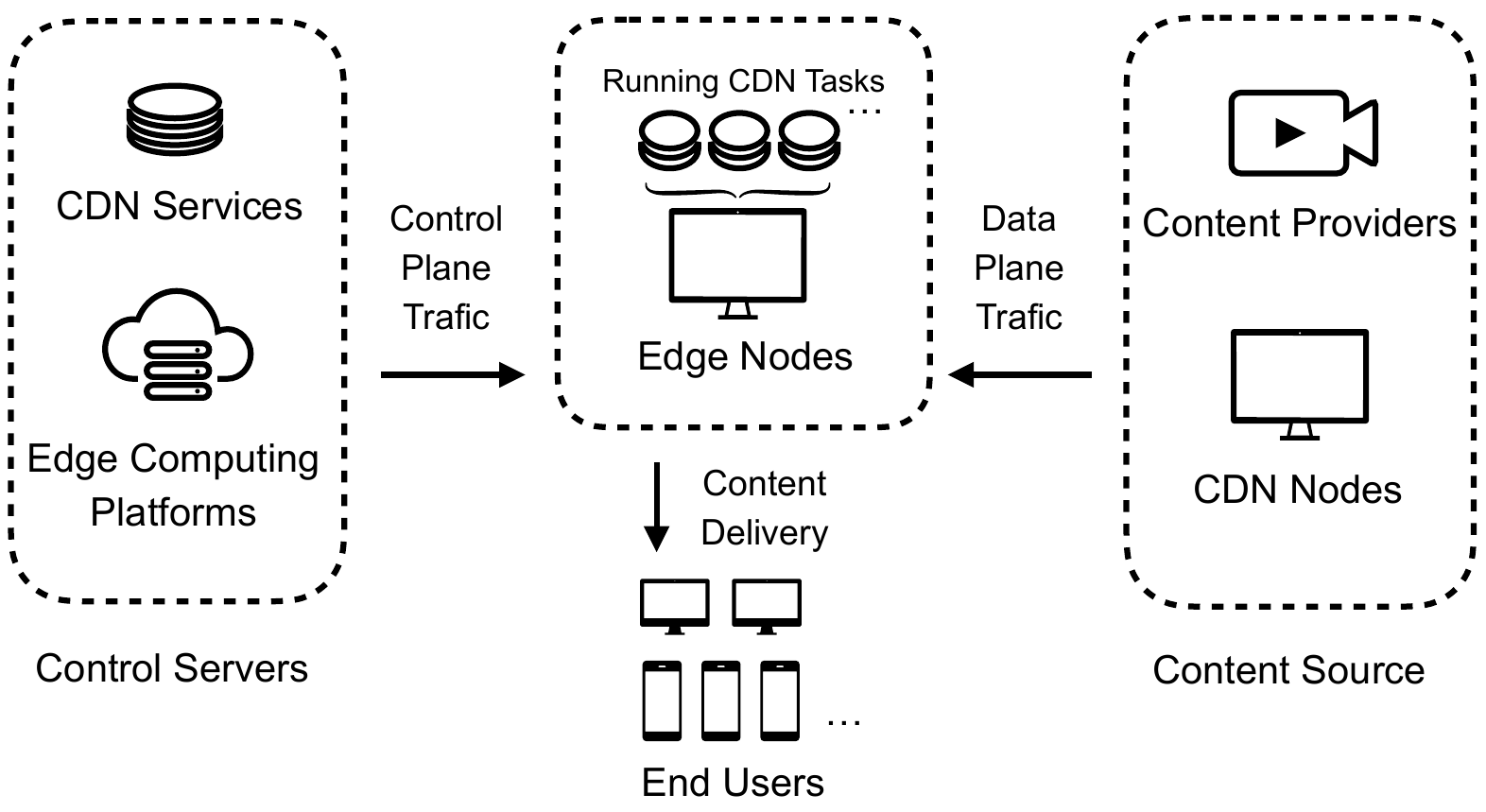}
    \caption{The open edge computing ecosystem.}
    \label{fig:ecp_ecosystem}
\end{figure}

Leveraging aforementioned methodology, we have gained a deep understanding for the ecosystem of open edge computing. As shown in Figure~\ref{fig:ecp_ecosystem}, multiple parties have participated in this ecosystem. First of all, open edge computing platforms (OECPs) build up the pool of edge nodes through recruiting third-party devices, e.g, personal computers, mobile phones,  WIFI hotspots. As the return, these device owners, i.e., edge node operators, can earn a revenue that is typically proportional to the volume of resource consumption, e.g., bandwidth, computing, and storage resources. Then, Given the computing/bandwidth resources available in these edge nodes, various content delivery services (CDNs) are the main consumers which deploy their content delivery modules on edge nodes and  transform these edge nodes into content delivery servers (i.e., CDN nodes). One more indirect participant are the content providers (CPs), e.g., video streaming platforms, or cloud storage services. Instead of directly collaborating with OECPs, these CPs subscribe to one or more CDN services, which in turn, instruct edge nodes via the on-edge CDN modules to pull content from content providers, cache content locally, and deliver the content to end users on demand. Below, we provide more detailed characterization for these participants with regards to their scale, distribution, evolution, etc. 
And the resulting understanding of this ecosystem serves as an important prior before we can well profile the impact of the inherent security risks (\S\ref{sec:risks}).
\subsection{Open Edge Computing Platforms}
Regarding the two OECPs under our study, we focus on their technical mechanisms, which are observed through running their edge nodes, analyzing the edge traffic, and reading through their documentations. 

\subject{The pricing policy.} Regarding how much revenue an edge node can earn, both ECPs adopts a complicated but secret pricing model which takes into consideration multiple factors encompassing the device type, operation time, the packet loss rate, bandwidth consumption, etc.\cite{wxy_revenue,tt_revenue}. For instance, 
OneThingCloud promotes that, a typical edge node with 50Mbps upstream bandwidth can earn around 75RMB per month\cite{wxy_revenue}. On the other hand, none of the OECPs provide any details regarding their pricing policies for edge computing tenants. 

\subject{Technical mechanisms.}
In a nutshell, a typical edge node is responsible to orchestrate and monitor edge computing tasks. Once deployed and started, the edge node communicates with the OECP control server, which in turn, returns a list of edge computing tasks along with the instructions for deploying these tasks. Then, to deploy an edge computing task, the edge node downloads the necessary executable payloads and jump-starts the respective edge computing task. When executing edge computing tasks,  the edge node periodically sends logs (e.g., resource consumption statistics) to the control server for diagnosis and billing. Such a paradigm is applicable to edge nodes of both OECPs. One thing to note, edge computing tasks deployed on the same edge node can belong to different real-world parties, most of which are independent from and collaborate with the OECPs. However, the two OECPs differ in how they isolate these co-located edge computing tasks. Briefly,  a OneThingCloud edge node runs edge computing tasks as separate containers, whereas a TipTime edge node runs tasks under the root user and enforces no confinement for them, which obviously incurs non-negligible security concerns, as further elaborated in \S\ref{sec:risks}.

\subject{Edge computing tasks.} 
Regarding the types of edge computing tasks, all tasks observed in our deployment are CDN tasks. One explanation is that peer-to-peer CDN (PCDN) is in great demand on the China CDN market. However, variance still exists in their task operator and content types.
For instance, TipTime edge nodes under our deployment have been used to execute content delivery tasks from two CDN services, namely, YunFan CDN and  Wangsu CDN. Also, the CDN modules of YunFan have been observed to deliver content payloads for 7 different and highly popular content providers, e.g., Kuishou and Douyin in the domain of short video streaming. More details about CDN services and content providers will be elaborated in \S\ref{subsec:ecosystem_cdn}.

\subsection{Edge Computing Nodes}
Regarding edge computing nodes (i.e., edge nodes), we aim to profile their scale and distribution. However, the pool of edge nodes is a black box, to which only the respective edge platform has full access. 
Through analyzing the edge traffic, several side channels have been successfully identified, which allow us to gain either a lower-bound estimate or an upper-bound approximation for edge nodes of aforementioned two OECPs. 

\begin{table}
    \centering
    \footnotesize
    \caption{The stats of edge nodes as observed in edge traffic.}
    \label{tab:edge_nodes_traffic}
    \begin{tabular}{ccccc}
        \toprule
        Platform & Node Source &Node IPs & /8 IPv4 & ASes\\
        \midrule
        TipTime & YunFan CDN &  17,585 & 51 & 46\\
         OneThingCloud & Bilibili CDN &  1,817 & 49& 38\\
         OneThingCloud & Xingyu CDN & 2,818 & 32 & 5 \\
         Both & All & 22,214 &       54 &  67    \\
          % TipTime pDNS & 28,212,518& 90 & -&- \\
         \bottomrule
    \end{tabular}
\end{table}

\begin{figure}
    \centering
    \includegraphics[width=.8\columnwidth]{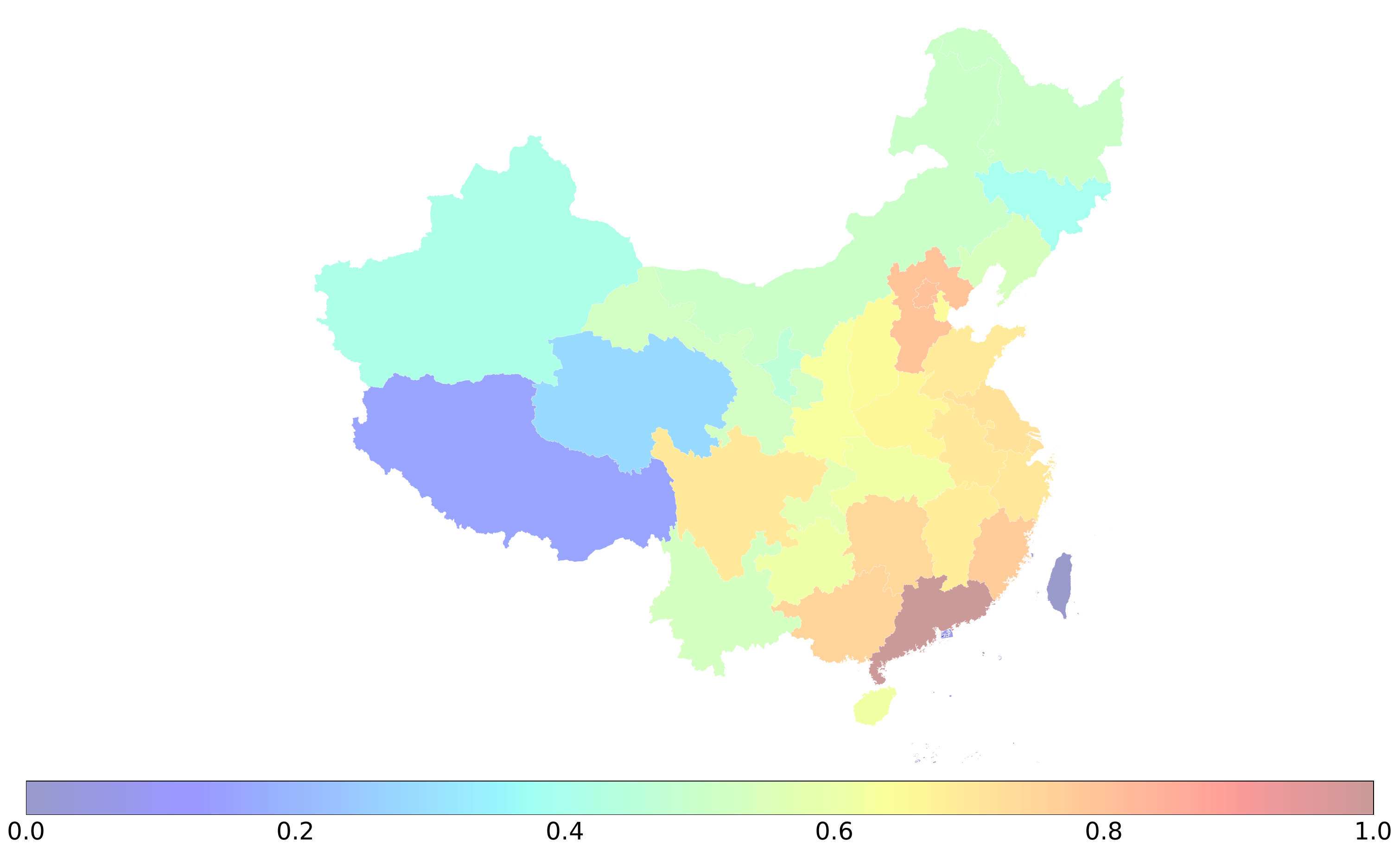}
    \caption{The geographical distribution of edge nodes observed in edge traffic.}
    \label{fig:heatmap_for_edge_node_in_china}
\end{figure}

\subject{Edge nodes observed in edge traffic.} Since edge nodes are mainly used as CDN nodes, counterpart edge nodes will communicate with ones under our control to either pull or push content payloads. 
Therefore, the first lower-bound estimator is the edge nodes observed in edge traffic. 
For TipTime, we observe that the CDN nodes of YunFan can serve as a good proxy for TipTime edge nodes. This is because there is a close collaboration between TipTime and YunFan. And a  TipTime edge node will activate the YunFan CDN task by default as long as it is started.  
Then, leveraging traffic signatures of YunFan CDN nodes, IPs of YunFan CDN nodes can be extracted from the edge traffic of TipTime. When it comes to OneThingCloud, the underlying operator of OneThingCloud provides its own CDN service named Xingyu CDN, 
and a typical OneThingCloud edge node will by default enable the Xingyu CDN module. Therefore, the CDN nodes of Xingyu can serve as a good proxy for edge nodes of OneThingCloud. Furthermore, we have also identified another proxy for OneThingCloud edge nodes. Specifically, Bilibili CDN deploys its CDN module on edge nodes of OneThingCloud. And when a remote Bilibili CDN node initiates communication with our edge nodes, it would clearly mention in the plaintext hello message regarding its device type and suggest whether it is another OneThingCloud edge node or not. The same case was observed when our edge nodes initiated Bilibili traffic flows with remote Bilibili CDN counterparts.  Therefore, remote Bilibili CDN nodes that are self-claimed as OneThingCloud nodes can also be used to lower-estimate the scale and distribution of the edge nodes for the OneThingCloud platform. 

Given these observations as well as the traffic signatures of respective traffic flows (e.g., flows between remote YunFan CDN nodes and our edge nodes) , we are allowed to quantitatively identify edge nodes that have ever communicated with ones under our control. 
Specifically, as learned from edge traffic, 22,214 edge node IPs have ever communicated with ones under our control, among which, 17,585 are YunFan CDN nodes (Tiptime edge nodes), and 2,818 are Xingyu CDN nodes (OneThingCloud nodes), and 1,817 are Bilibili CDN nodes that claim to be OneThingCloud nodes.   As listed in Table~\ref{tab:edge_nodes_traffic}, despite all located in China, they are widely distributed across 54 /8 IPv4 network blocks, 67 different autonomous systems, and most provinces in China (Figure~\ref{fig:heatmap_for_edge_node_in_china} ). One thing to note, we believe edge nodes observed in traffic can only serve as a lower-bound estimate, as not all available edge nodes would communicate with ones under our control, which is further demonstrated by the edge nodes observed in passive DNS.

\finding{
Both OECPs emerged in early 2021 and have millions of daily active edge nodes that are widely distributed across network blocks.
}

\begin{table}
    \footnotesize
    \centering
        \caption{The domain patterns for edge nodes.}
    \label{tab:edge_nodes_pattern}
    \scalebox{0.8}{
    \begin{threeparttable}
    \begin{tabular}{cccccc}
        \toprule
    Platform & CDN & Node Domain Pattern \\
            \midrule
            \multirow{6}{*}{TipTime} & \multirow{6}{*}{YunFan} 
            & *yf*.nodeedge.cn\\
            & & *yf*.free-lbv6.idouyinvod.com\\
            & & *yf*.dl.jcloudimg.com \\
            & & ndsv1-*.jdcloudstatus.net /*yf*.jdcloudstatus.net\\
            & & ndsv1-*.cdnnode.cn\\
            & & ndsv1-*.cachenode.cn\\
            OneThingCloud & Xingyu & *.vld.szbdyd.com /*.v2l.szbdyd.com\\
         \bottomrule
    \end{tabular}
    \end{threeparttable}}
\end{table}

\begin{table}
    \footnotesize
    \centering
        \caption{The stats of edge nodes as observed in passive DNS.}
    \label{tab:edge_nodes_dns}
    \scalebox{0.8}{
    \begin{threeparttable}
    \begin{tabular}{cccccc}
        \toprule
    Platform & Node FQDNs &Node IPs~\tnote{1} & IPv6 & /8 IPv4 & ASes~\tnote{2}\\
        \midrule
        TipTime  & 4,233,571,373 & 28,212,313 &9,416,567 & 89& 114\\
         OneThingCloud & 100,492,251  & 7,383,677 &  4,654,242& 255& 182 \\
         Both & 4,334,063,624  & 34,364,400&14,070,775 &255 & 237\\
          % TipTime pDNS & 28,212,518& 90 & -&- \\
         \bottomrule
    \end{tabular}
        \begin{tablenotes}
        \item [1] Both IPv4 and IPv6 addresses.
        \item[2] Each platform has 500K IPs sampled to query IPinfo for autonomous systems (ASes). 
    \end{tablenotes}
    \end{threeparttable}}
\end{table}

\subject{Edge nodes observed in passive DNS (pDNS).}  Furthermore, we observe that  YunFan CDN assigns to each CDN node unique FQDNs (fully qualified domain names) and such FQDNs follow unified subdomain patterns ( see Table~\ref{tab:edge_nodes_pattern}). Therefore, querying passive DNS with these FQDN patterns can reveal historically active CDN node IPs, which provides another channel to upper-bound estimated edge nodes of TipTime. The same case also applies to the pair of OneThingCloud and Xingyu CDN.  
To further profile the edge nodes, we further queried a representative passive DNS dataset from our industry collaborator  with the FQDN patterns belonging to YunFan and Xingyu. When querying the passive DNS, a historic DNS record will be returned only if it matches one of aforementioned CDN domain patterns and it used to be active during the time period between January 2021 and November 2023. We consider this time window because the number of edge nodes that were active before 2021 is negligible. 
As listed in Table~\ref{tab:edge_nodes_dns}, compared with edge nodes observed in edge traffic, edge nodes cumulatively recorded in pDNS have a much larger scale. Particularly, 28 million edge node IPs (4 billion FQDNs) have been observed for TipTime, among which, a randomly sampled subset of 500K IPs reveal a wide distribution in  114 autonomous
systems and 12 different countries. Still, most edge node IPs of TipTime (99.93\%) are located in China, which is followed by Japan (0.03\%) and Singapore (0.02\%). 

\begin{figure}
    \centering
    \subfigure[
        Daily active edge nodes.
        \label{fig:daily_active_edge_nodes}
    ]{
        \includegraphics[width=.45\columnwidth]{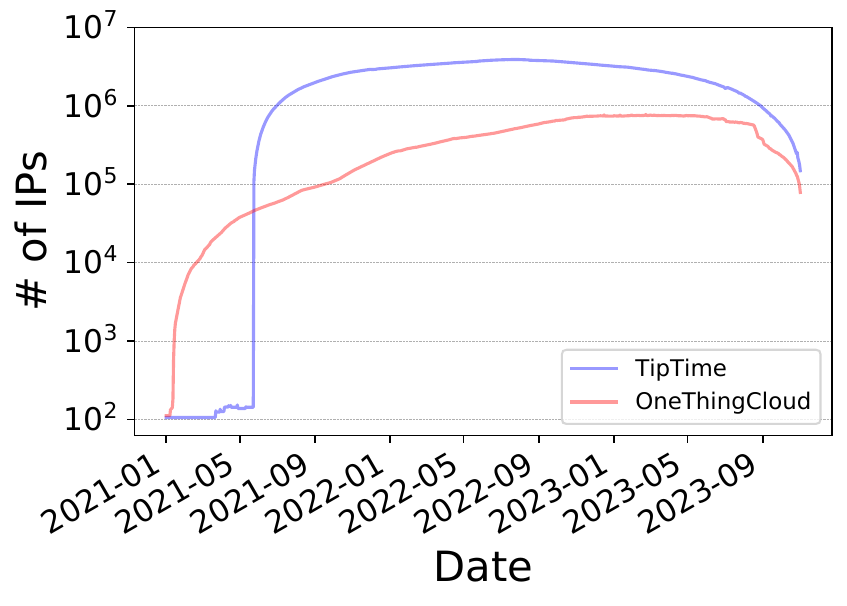}
    }
    \hfill
    \subfigure[
        Daily new edge nodes.
        \label{fig:daily_new_edge_nodes}
        ]{
        \includegraphics[width=.45\columnwidth]{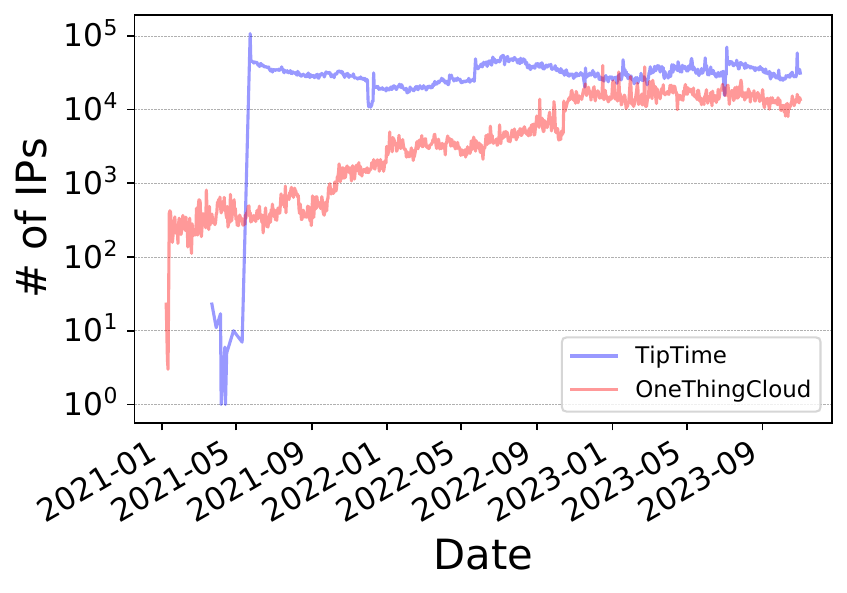}
    }
    \caption{The temporal evolution of daily active edge nodes.}
    \label{fig:daily_evolution}
\end{figure}

\subsubject{The temporal evolution of edge nodes.} Leveraging the timestamps that exist in each passive DNS record, we are allowed to profile not only the temporal evolution of edge nodes but also the lifetime of each individual edge node. 
Figure~\ref{fig:daily_active_edge_nodes} presents how the number of daily active edge nodes evolves across the time period between January 2021 and November 2023, while Figure~\ref{fig:daily_new_edge_nodes} presents the number of new edge nodes that daily emerge for the same period.
As we can see that both OECPs emerged in early 2021 with a small scale of edge nodes, and then quickly ramped up their pools of edge nodes. 
And the largest volume of daily active edge nodes for TipTime is 3.9 millions while it is  0.8 million for OneThingCloud. On the other hand, we can also observe a large volume of newly emerging edge node IPs, as shown in Figure~\ref{fig:daily_new_edge_nodes}. For instance, TipTime has over 10k new edge node IPs observed every day between June, 2021 and November, 2023. Considering the scale of edge node IPs is stable since late 2021, the continuous emergence of new edge node IPs suggests a non-negligible churning rate. We also profile the lifetime of an edge node. We observe that  most edge nodes of both
OECPs have a short
lifetime, which is very different from traditional cloud servers. For
instance, over 70\% TipTime edge nodes have the lifetime shorter
than 10 days while it is 76\% for OnethingCloud edge nodes. For more details, please refer to Appendix~\ref{appendix:edge_nodes}.

\subsection{Edge-Assisted Content Delivery}
\label{subsec:ecosystem_cdn}

\subject{CDNs and content providers.} As aforementioned,  
all edge computing tasks observed in our study are content delivery tasks which involve the collaboration between CDN services and the open edge computing platforms. 
In total, we have observed the tasks of 6 different CDN services, among which, 2 are observed on the TipTime platform and 4 are on the OneThingCloud. Also, through analyzing the traffic flows of these content delivery tasks, we have identified 16 upstream content providers that subscribe to one or more of these 6 CDN services and have their content payloads delivered through edge nodes of the two OECPs. 
For the full list of CDN services along with the respective upstream content providers, please refer to Table~\ref{tab:cdns} in Appendix~\ref{appendix:content_delivery}.
Also, all CDN services except for Xingyu CDN are independent from the two OECPs.

\subject{CDN nodes observed in edge traffic.}
Also, as detailed in \S\ref{sec:method}, most traffic flows of CDN tasks show distinguishable features for deciding whether a remote party is a CDN node or not. Leveraging these distinguishing features, we are allowed to profile the scale and distribution of CDN nodes that have ever communicated with edge nodes under our control.  Despite being a lower-bound estimate,  these CDN nodes turn out to be large-scale and widely distributed, regardless of the CDN services. In total, we have observed over 125K distinct CDN node IPs which feature a wide distribution in 61 /8 IPv4 network blocks and 120 different ASes. For more details, please refer to Appendix~\ref{appendix:cdn_nodes_in_edge_traffic}.

\subject{Content payloads under delivery.} Across both OECPs, edge nodes are used to deliver content of diverse categories, which include not only traditional static web files (Javascript and media files), but also emerging content types (program files and machine learning models). Particularly, edge nodes under our control used to deliver files of machine learning models to end users,  e.g., .tflite, .model, .mlmodel, and .weights. Further investigation reveals that they are likely used for face-relevant processing for users of Kuaishou. For more details, please refer to Appendix~\ref{subsec:content_under_delivery}.

In summary, we conclude that edge nodes of OECPs play an important role in Internet-wide content delivery, which motivates a strict vetting of their potential security and privacy risks.

\section{The Security Risks}
\label{sec:risks}
In this study, we have identified and profiled, for the first time, a set of concerning security vulnerabilities for two representative open edge computing platforms (OECPs). Although some of the identified security risks should be attributed to implementation errors,  most are inherent in the design of OECPs, such as the low threat reputation of byzantine or even malicious edge nodes, the privacy risks against end users, and the exposure of long-term credentials (e.g., TLS private keys) to potential attackers, etc. 

\subject{The threat model.} Before diving into security risks, it is necessary to first highlight the threat model upon which we reason about the feasibility and impact of any security risk. First of all, 
as edge nodes are recruited from the public along with a low vetting bar, which only exists in the OECP paradigms, we assume some edge nodes can be byzantine or malicious, i.e., under the full control of an attacker. Then, when an edge node under the attacker's control serves as a CDN node, we assume the attacker has full access to the content payloads under delivery. Although many content payloads can be either encrypted or proprietorially encoded when getting stored in local disks, we observe that such operations of encoding or encryption are conducted on the edge side and are thus visible to the edge node. However, although an attacker can  modify the content payloads that are delivered through the edge node under its control, we argue that such a modification can be easily detected and the modified payloads can be discarded.  This is because CDN services and content providers can easily instruct the end-user devices to conduct integrity checking for content payloads received from edge nodes and the materials of integrity checking (e.g., signatures, or messaging authentication code) can be transmitted through other secure channels, e.g., a direct secure connection between the end-user device and the server of the content provider. 

\subsection{The Security Testbed}
\label{subsec:method_security_testbed}
We first design and implement a security testbed so as to experiment potential security and privacy risks while avoiding any ethical concerns. 
As detailed in the threat model, some edge nodes can be under full control of an attacker. The main purpose of this testbed is thus to evaluate various man-in-the-middle attacks against edge traffic. To achieve this, this testbed deploys a mitmproxy~\footnote{https://mitmproxy.org}  on the same host of the edge node under evaluation. Then, to carry out a MITM experiment, a straightforward method is to redirect all traffic flows to the mitmproxy node for traffic interception, which however, would incur ethical concerns. This is because edge nodes are mainly used for content delivery towards end users, and any MITM attempts can disrupt real-world content delivery towards real-world users. To avoid this issue, our MITM experiments are limited to traffic flows not directly relevant to content delivery. To achieve this, utilizing the signatures to distinguish different traffic flows,  the host-wide firewall is carefully configured so that only traffic flows under our consideration will be redirected to the mitmproxy node while others relevant to content delivery will not be impacted. 

Then, when a TLS traffic flow was vulnerable to MITM attacks, the payload would be exposed to the mitmproxy node and also to our researchers, which however is still ethically concerning. To address this issue, a script is designed to run as a mitmproxy addon and automatically verifies if a MITM attack is successful or not without revealing the intercepted plaintext traffic to our researchers. Therefore, during our MITM attack experiments, only TLS traffic flows that are not directly relevant to content delivery will be experimented. Also, even if a traffic flow is successfully intercepted, none of its payload bytes will be exposed to our researchers, nor does our experiment script manipulate the intercepted traffic. 

Due to aforementioned ethical considerations, our MITM experiments have a limited coverage of edge traffic flows, i.e., only the edge-wide control-plane traffic and the CDN-specific control-plane traffic are tested. Also, our MITM experiments consider only the most common but simple scenario wherein the attacker sitting at an intermediate hop utilizes a self-signed root certificate to dynamically issue leaf certificates for the requested domain name. Despite these limitations, we are surprised to observe that edge nodes and CDN modules across the two OECPs fail to validate the TLS certificate for many critical traffic flows, for which, more details will be presented in \S\ref{subsec:security_implementation_errors}.

\subsection{The Exposure of Credentials to Attackers}
% \subsection{The Exposure of Long-Term and Cross-Edge-Node Credentials to Attackers}
\label{subsec:security_protocol}
\finding{We find that edge nodes across platforms tend to share and locally store long-term TLS credentials, which renders a non-negligible MITM attacking surface for TLS traffic of content delivery. }

One fundamental security vulnerability resides in the credential management of edge nodes. As edge nodes can be operated by any third parties including the attackers, any credentials (e.g., TLS certificates and private keys) transmitted to and stored at the edge node can be theoretically compromised. However, on the other hand, it is necessary for edge nodes to locally store and make use of various credentials that are long-term (non-ephemeral). For instance,  when edge nodes are used as content delivery servers, they need to store multiple pairs of TLS certificates and private keys. Otherwise, when end-user TLS clients request content from these edge nodes, they will not be able to authenticate these edge nodes as the servers of respective domain names. To summarize, the necessity of storing long-term credentials and the existence of byzantine (malicious) edge nodes creates a security tension that is inherent in open edge computing platforms.

\begin{figure}
    \centering
    \includegraphics[width=.7\columnwidth]{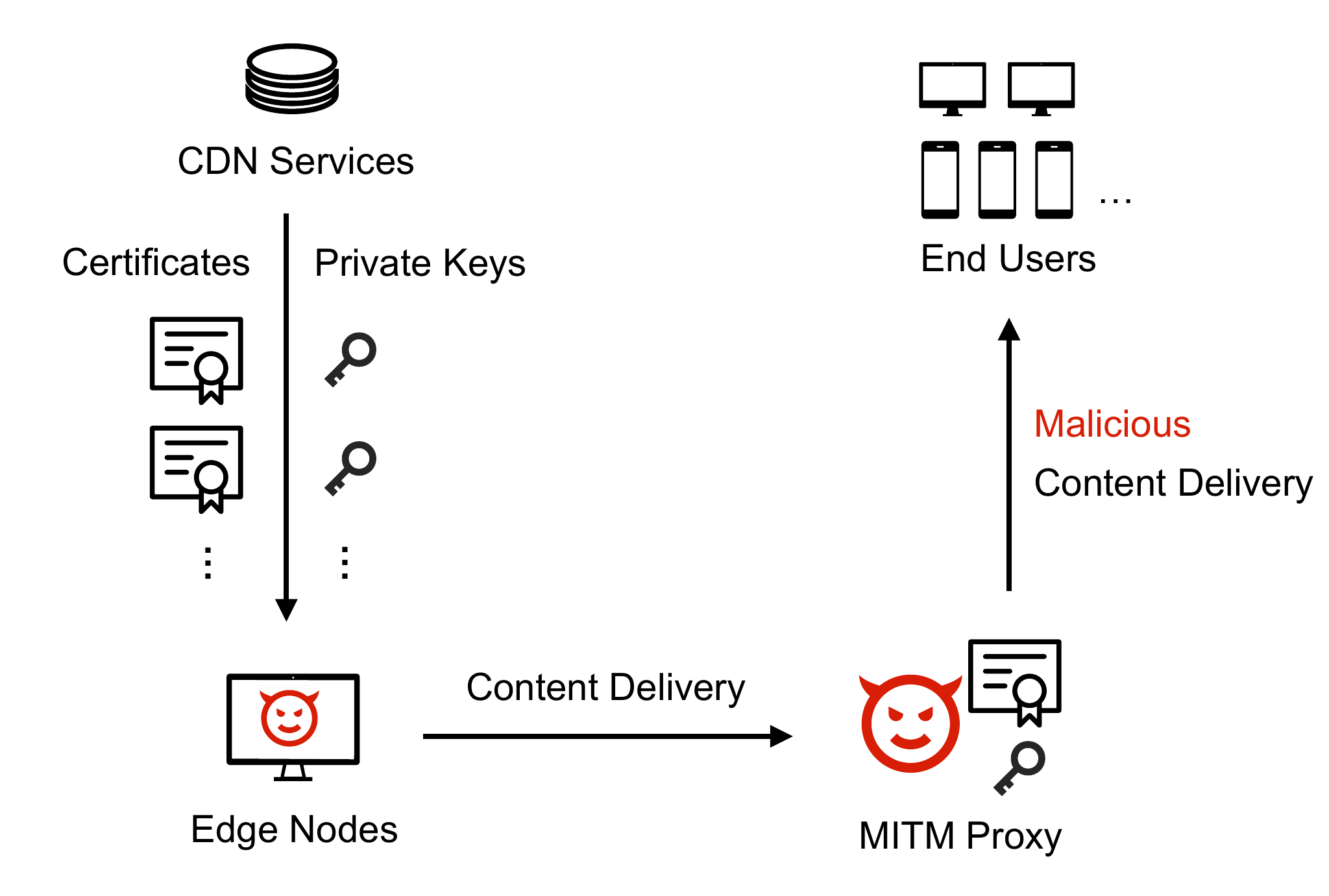}
    \caption{The scenario of the man-in-the-middle attacks. Once edge nodes operated by the attacker gain access to TLS credentials, it could control the content delivery flow.} 
    \label{fig:man-in-the-middle}
\end{figure}
This tension has led to a severe vulnerability for edge nodes of TipTime. Specifically, the YunFan CDN module, the default computing task of TipTime edge nodes, delivers content through TLS to end users of multiple content providers,  to achieve which it needs to act as TLS servers of the respective content providers. As observed in edge traffic, the YunFan CDN module has acted as the TLS servers for 6 different domain names of 3 distinct content providers, e.g., \textit{*.free-lbv6.idouyinvod.com} for Douyin. To act as a legal TLS server, the CDN module and the respective edge nodes must be assigned with a valid TLS certificate as well as the respective private key. However, it turns out that the management of these TLS private keys have multiple concerning issues. The first issue is that these TLS certificates are issued with a long lifetime, which ranges from 365 days (i.e., the certificate for domain ndsv1-*.cachenode.cn) to 396 days. 
As we can see from Figure~\ref{fig:man-in-the-middle}, the attacker can have full control of the edge nodes as well as the TLS credentials.
Once the attacker had got access to such a pair of the certificate and the private key, it could conduct transparent MITM attacks at least for the respective edge node and thus break the confidentiality for the content delivery flows between the edge node and the end users. In addition, it may also break the integrity of the content under delivery unless integrity checking metadata (e.g., message authentication code) is delivered to the end user (the TLS client) through other secure channels and the end user has conducted extra integrity checking for the received content.  

Then,  the second issue is that all edge nodes running the YunFan CDN module turn out to \textit{share the same set of pairs of the TLS certificates and private keys}. For instance, when the 3 TipTime edge nodes under our control delivered content to end users of Douyin, they used the same TLS certificate of which the subject name is \textit{*.free-lbv6.idouyinvod.com} and the certificate lifetime is from October 29, 2021 to November 30, 2022. The sharing of the same TLS credentials across edge nodes may effectively lower the management cost but at the significant cost of security. In such a case, as long as one of these edge nodes got compromised and had its TLS private key stolen by the attacker, all edge nodes and their content delivery TLS traffic would be subject to MITM attacks, which could be further exaggerated by the long lifetime of the respective TLS credentials. 

One more issue resides in the distribution of these credential materials. Specifically,  when the YunFan CDN module is instantiated on an edge node, it will be instructed to download the TLS credentials from the control server. Although the payload is transmitted over HTTPS, the credential hosting server has enforced no authentication for downloading requests and the  downloading URLs have also got leaked in the plaintext logging traffic towards the control server, which allows any intermediate hops to learn these downloading URLs from passively observing the plaintext logging traffic and then download the TLS credentials even without direct control of a running edge node. We first observed this issue in late 2022, which had then been fixed by YunFan in early 2023. However, due to the failure of TLS certificate verification as detailed below in \S\ref{subsec:security_implementation_errors}, an attacker can still learn the credential downloading URLs even if it has no direct access to any edge nodes.

For all the issues and impact above, we discuss the solutions in Appendix~\ref{appendix:security_recommendations}, e.g, adopt the short-lived and node-specific credentials.

\subsection{Low Threat Reputation of Edge Node IPs}
\label{subsec:threat_node_ip}
As an edge node can be recruited from any third party that is \textit{untrusted} or even malicious, they can be co-located on the same IP address with various malicious programs or activities. Such a co-location with malicious activities can lead to a low threat reputation for edge node IPs, which in turn can trigger detection alarms or even network traffic blocking for the legitimate edge computing activities (e.g., content delivery). On the other hand, as these edge platforms are open to any third parties to join as edge nodes and earn revenue, miscreants (e.g., botnet operators) may abuse these platforms as a monetization channel for compromised devices under their control,  just like what have been observed in cryptojacking and residential proxies~\cite{carlin2019you, tekiner2021sok, mi2019resident, mi2021your}. 

\begin{table}
    \centering
    \footnotesize
    \caption{The threat reputation of edge node IPs as revealed by the proprietary threat platform. 
    }
    \label{tab:stats_private_threat}
    \scalebox{0.8}{
    \begin{threeparttable}[online]
        
    \begin{tabular}{llcccc}
        \toprule
        Edge  & Group  & w MTFs\tnote{1} & $\ge 5$ MTFs \tnote{1} & $\ge 10$ MTFs~\tnote{1} & Median MTFs~\tnote{2}\\
        \midrule
        \multirow{4}{*}{TipTime} 
        & $\text{Edge}_{\text{Traffic}}$ & 88.06\% &    83.09\% &     79.66\% & 359 \\
        & $\text{CDN}_{\text{Traffic}}$ &83.44\% &    78.67\% &     75.45\% & 363 \\
        & $\text{Edge}_{\text{DNS}}$ &62.72\% &    58.09\% &     55.37\% & 283 \\
        & All &   65.94\% &    61.30\% &     58.50\% &          297 \\
        \hline
        \multirow{4}{*}{OneThing}
        & $\text{Edge}_{\text{Traffic}}$ & 57.07\% &    49.60\% &     45.97\% &          110 \\ 
        & $\text{CDN}_{\text{Traffic}}$ & 78.57\% &    71.71\% &     67.89\% & 202 \\
        & $\text{Edge}_{\text{DNS}}$ & 29.70\% &    26.80\% &     25.13\% & 173 \\
        & All & 64.41\% &    58.52\% &     55.26\% &          196  \\
        \hline
         \multirow{4}{*}{Both}
         & $\text{Edge}_{\text{Traffic}}$ & 78.25\% &    72.48\% &     68.99\% &          280 \\
         & $\text{CDN}_{\text{Traffic}}$ &80.11\% &    73.91\% &     70.28\% &          244\\
         & $\text{Edge}_{\text{DNS}}$ &58.00\% &    53.62\% &     51.05\% & 273\\
         & All & 65.42\% &    60.36\% &     57.41\% &          260 \\
        \bottomrule
    \end{tabular}
    \begin{tablenotes}
        \item [1] The fraction of IPs with one or more malicious traffic flows (MTFs) observed during January, 2022 and November, 2023.
        \item [2] The median number of MTFs for node IP addresses that have one or more MTFs. 
    \end{tablenotes}
   \end{threeparttable}}
\end{table}

We thus move to quantitatively profile the threat reputation of edge node IPs, which is achieved through analyzing threat reports aggregated by two representative threat intelligence platforms. 
One is a proprietary threat intelligence platform maintained by a top security vendor, considering most edge nodes are located in China. Then the global and publicly available VirusTotal~\cite{vt_api} is also considered.    VirusTotal focuses more on how an IP acts as a server in malicious activities, such as hosting malware or phishing websites. While the private threat platform is very helpful to profile how an IP serves as a malicious client when involving malicious activities, e.g., contacting a botnet C2 server.
Then, due to rate limits enforced by both platforms as well as the large scale of observed edge nodes, a random sampling strategy was applied to edge node IPs when querying both platforms. As discussed in \S\ref{sec:ecosystem}, the captured edge node IPs belong to different groups, depending on the respective edge platform, whether they were captured from the raw traffic or the passive DNS, and whether they have been confirmed to be edge nodes or just CDN nodes (i.e., potential edge nodes). For each OECP, we define three groups of node IPs. One is the group of edge node IPs as observed in edge traffic, which we name as $\text{Edge}_{\text{Traffic}}$. Then, it is $\text{CDN}_{\text{Traffic}}$ which encompasses IP addresses of CDN nodes as observed in edge traffic. Lastly, the third group $\text{Edge}_{\text{DNS}}$ consists of edge node IP addresses as learned from passive DNS. Then, to query both threat intelligence platforms, we randomly sampled up to 10K IPs from each of all these groups for each OECP.

\subject{Threat reports from the proprietary threat intelligence platform.} We first look into malicious traces of edge nodes as learned from the proprietary threat intelligence platform, which reveals that edge node IPs are concurrently involved in malicious activities that feature both a large scale and diverse categories. 
As shown in Table~\ref{tab:stats_private_threat}, a large portion of edge node IPs have been associated with malicious traffic flows (MTFs) that were captured by the proprietary threat dataset between January, 2022 and November, 2023. Particularly, among TipTime edge nodes directly observed in edge traffic, 79.66\% have been involved in 10 or more MTFS, while it is 45.97\% for OneThingCloud edge nodes. Also, among edge node IPs associated with one or more MTFs, the median number of MTFs is 275, while the average one is 34,227.  Even if we increase the bar for associating an IP with MTFs through counting only MTFs that are captured on the exact date when an IP serves as an edge node, we can still observe such an extensive involvement in malicious activities. While only counting MTFs captured for dates when an IP serves as an edge node, 6.15\% TipTime edge nodes are still associated with one or more MTFs while it is 2.94\% for OneThingCloud edge nodes. For more details, please refer to Table~\ref{tab:stats_private_threat_same_date} in Appendix~\ref{appendix:threat_reputation}.

\begin{table}
    \centering
    \footnotesize
    \caption{Top 5 categories of malicious traffic flows.}
    \label{tab:top_class_mtf}
    \scalebox{0.93}{
    \begin{threeparttable}
        \begin{tabular}{
    >{\centering\arraybackslash}p{0.33\linewidth}
    >{\centering\arraybackslash}p{0.16\linewidth}
    >{\centering\arraybackslash}p{0.12\linewidth}
    >{\centering\arraybackslash}p{0.1\linewidth}
    >{\centering\arraybackslash}p{0.1\linewidth}}
            \toprule
              Category &      MTFs &  \% MTFs &  \% Edge IPs  & \% CDN IPs\\
             \midrule
             Botnet & 1.37B	& 68.92\%&11.08\%&11.84\% \\
            RAT~\tnote{1} & 312M&15.69\%&55.90\%&59.59\%  \\
            Illicit promotion & 111M &5.60\%&48.88\%&50.73\% \\
            Cryptojacking & 67M & 3.38\%&17.17\% &18.09\%  \\
            Malicious downloads & 44M&2.21\%&4.92\%&5.19\%\\
    \bottomrule
        \end{tabular}
    \begin{tablenotes}
        \item [1] RAT stands for the remote access trojan.
    \end{tablenotes}
    \end{threeparttable}}
\end{table}
We then take a closer look into categories of the edge-involved MTFs. In total, the proprietary threat platform has observed 13 different categories of MTFs. Table~\ref{tab:top_class_mtf} presents top 5 along with their contribution to MTFs and the involved edge nodes. These top five categories include botnet, remote access trojan (RAT), illicit promotion,  cryptojacking, and malicious downloads. Particularly, over 1.3 billion botnet traffic flows have been captured, which involve 11\% of all the sampled edge node IPs. On the other hand, 55.90\% edge node IPs are involved in MTFs of RAT which suggest that one or more machines attached to these IPs are compromised with RATs installed. For instance, 48\% edge node IPs have involved in traffic flows towards the C2 server (i.e., \textit{pro.csocools.com}) of DoubleGuns, a RAT campaign being active in the last few years~\cite{shuangjiang}.

We also case study edge node IPs that are associated with the largest number of MTFs. We observe that these top cases are associated with MTFs that are not only diverse in malicious categories but also span a long period. For instance, \textit{58.221.114.86}, is a TipTime edge node used to deliver content payloads to edge nodes under our control. It has been involved in 6 different categories of over 1.2 billion MTFs. Also, starting from July, 2022, this edge node IP continuously has MTFs captured for a total of 357 days by November, 2023. To further investigate why this IP can generate so many MTFs, we looked into the detailed MTFs and found out that the majority of MTFs are DNS queries towards \textit{qq603535.3322.org}, a botnet C2 server. Other malicious domains contacted by this IP include \textit{testjj.com}, \textit{niria.biz}, etc.
One more example is \textit{121.10.143.23}, another TipTime edge node IP that used to deliver multiple content payloads to edge nodes under our control. However, this edge node has MTFs captured for
646 days between 01/25/2022 and 11/14/2023.
Furthermore, \textit{222.173.104.238} used to be a OneThingCloud edge node and it has involved 12 different categories of 123 million MTFs that are continuously observed for 647 days.

When it comes to VirusTotal, a non-negligible fraction of edge node IPs are considered as malicious  for either hosting malicious URLs or distributing malware payloads. For more details, please refer to Appendix~\ref{appendix:threat_reputation}. 
In summary, a large portion of edge node IPs are concurrently involved into diverse malicious activities, which can likely render a low threat reputation and thus disrupt the legitimate edge computing tasks. 
For example, the CDN domain cdn.thunderstore.io was blocked  because a malicious user had uploaded a crypto miner to this site~\cite{CDNblocked}.

\subsection{The Validation Failures of TLS Certificates}
\label{subsec:security_implementation_errors}
\begin{table}
    \centering
    \footnotesize
    \caption{The issue of certificate validation failures for edge nodes, wherein \ding{56} denotes validation failures. }
    \label{tab:tls_validation_edge_node}
    \begin{threeparttable}
        \begin{tabular}{cccc}
            \toprule
            \multirow{2}{*}{Edge Type} & \multicolumn{3}{c}{Traffic Category}\\ 
            & Control\tnote{1} & Logging\tnote{1} & Task Payload\tnote{1} \\
            \midrule
             TipTime& \ding{56} & \ding{56}& \ding{56} \\
             OneThingCloud & \ding{52} & \ding{56} & \ding{52}  \\
             \bottomrule
        \end{tabular}
        \begin{tablenotes}
            \item [1] Task Payloads denotes flows for downloading deployment payloads of edge computing tasks.
        \end{tablenotes}
    \end{threeparttable}
\end{table}
Aforementioned security risks are inherent in the design of the edge platforms, while the failure of validating TLS certificates should be attributed to either implementation errors or bad security practices. In a nutshell, we observe and demonstrate that edge nodes of both OECPs fail to verify the server TLS certificate for  part of the TLS traffic flows. Although it should be considered as an implementation error, we are surprised to see the significant extent it has shown up in edge computing traffic flows. 
Specifically, when connecting to upstream servers via TLS, some edge nodes and CDN modules fail to properly validate the certificates of many TLS servers, which allows a MITM attacker with no knowledge of the server's private key to break the confidentiality and integrity of the TLS traffic. Such a failure of certificate validation varies across edge platforms and CDN modules as well as the traffic categories (e.g., traffic for logging, and traffic for caching content). As listed in Table~\ref{tab:tls_validation_edge_node}, a TipTime edge node is subject to this vulnerability for all the TLS traffic towards upstream servers, while only the logging traffic of OneThingCloud edge nodes shares this vulnerability. However, one thing to note, our certificate validation test (\S\ref{subsec:method_security_testbed}) considers only a simple attacker that is equipped with a self-signed root certificate and can thus serve as a lower-bound estimator for the TLS security risks. In another word,  even if the edge node had passed our test, it may still be vulnerable to more complicated and more advanced MITM attacks, e.g.,  a complicated MITM attacker may instead use a certificate issued from a legitimate CA to domains under its control, in which case, the edge node would still be vulnerable to MITM attacks when it fails to verify the field of the common name. 

Then, when it comes to CDN modules, as listed in Table~\ref{tab:tls_validation_cdn}, the CDN TLS traffic towards upstream servers can be divided into four categories depending on the the purposes, namely,  control-plane communication, caching content from the upstream content servers, logging, and updating TLS credentials (e.g., certificate update). Among all the 6 CDN modules, 4 have adopted the standard TLS protocols for communication with the upstream servers, all of which suffer from the certificate invalidation risk for one or more categories of TLS traffic. Particularly, 
the YunFan CDN is vulnerable to certificate invalidation for all the four categories of TLS traffic. One thing to note, all these certificate invalidation results have been confirmed by our ethical MITM experiments.

\begin{table}
    \footnotesize
    \centering
    \caption{The issue of certificate validation failures for CDN modules, wherein \ding{56} denotes validation failures. }
    \label{tab:tls_validation_cdn}
    \begin{threeparttable}
        \begin{tabular}{ccccc}
            \toprule
            \multirow{2}{*}{CDN Task} & \multicolumn{4}{c}{Traffic Category}\\ 
            & Control\tnote{1} & Caching\tnote{1}&  Logging\tnote{1} &  Certificate Update\tnote{1} \\
            \midrule
             YunFan CDN & \ding{56} & \ding{56}& \ding{56}& \ding{56} \\
             Wangsu CDN & N/A\tnote{2} & N/A & N/A & N/A\\
             Xingyu CDN & \ding{56} & N/A & \ding{56}& N/A\\
             Bilibili CDN & \ding{52} & \ding{56} & \ding{56} & N/A\\
             Baidu CDN & N/A & N/A & N/A & N/A \\
             Xunlei CDN & N/A & \ding{56} &\ding{56}  & N/A\\
             \bottomrule
        \end{tabular}
        \begin{tablenotes}
            \item [1] 
            Caching denotes flows for caching content payloads, while Certificate Update refers to flows for retrieving TLS certificates and private keys.
             \item [2] N/A denotes the traffic flows are either  unavailable or they are not TLS.
        \end{tablenotes}
    \end{threeparttable}
\end{table}

\subsection{Privacy Risks Against End Users}
Edge-assisted content delivery is observed to  incur considerable privacy risks to end users, as elaborated below. 

\subject{Plaintext content delivery traffic}. One concerning privacy risk is that a large portion of content delivery traffic is delivered in plaintext.
Particularly, as observed from the edge traffic,  plaintext traffic accounts for 23.49\% content delivery traffic and involves 277K distinct recipient IP addresses as well as content payloads belonging to 5 different content providers. These results suggest a non-negligible privacy concern for end users. As the delivery of content via plaintext traffic flows allow any hop on the path to learn who is requesting what content. 

\subject{The information leakage to untrustworthy edge nodes.} We also observe that end users of edge-assisted content delivery are subject to information leakage to edge nodes. This is because a typical edge node knows not only the end-user IP addresses but also the plaintext content payloads under delivery. As an edge node in OECP can be malicious, such leakage can enable a potential attacker to link an end-user IP address with the content payloads delivered to it, which can provide a side channel for miscreants for targeted censorship and privacy infringement. For instance, a miscreant can run edge nodes and monitor what videos the nearby residents are watching. Furthermore, as learned from quantitative analysis, a typical edge node tend to deliver content to thousands of end-user IPs on a daily basis and these end-user IPs can be as far as thousands of kilometers from the edge node. What is also observed that the daily content delivery volume of a typical edge node could exceed 15GB. These quantitative results further highlight the concerning and practical impact of this information leakage.

\subject{The potential lack of user consent.} Given such a large scale of edge nodes observed, it is intuitive to wonder whether the respect  OECP has sought full authorization from the device owner when transforming a device into an edge node. As revealed by previous studies,  end-user devices of various types have been abused without sufficient user consent for suspicious activities such as residential web proxies~\cite{mi2019resident} and cryptojacking~\cite{carlin2019you}. And such kinds of suspicious activities have potentially paved new avenues for the monetization of botnets. However, we don't find any measures from the OECPs that can strictly vet the background of the untrustful edge node participants. In another word, as long as an executable payload can be deployed to a compromised device,  the miscreants can monetize the device by transforming it into an edge node. 

\subsection{Insufficient Confinement of OECP Tenants}
\label{subsec:confinement}
Another practical security issue is the insufficient confinement enforced by edge platforms for their tenants, e.g.,  co-located CDN modules. Similar to modern cloud computing, the two edge platforms allow the co-existence of multiple tenants (e.g., CDN modules) on the same physical device (the edge node). However, tenant confinement in edge-assisted content delivery turns out to be insufficient. Particularly, TipTime simply runs CDN modules with full privileges, which means a CDN module in TipTime has full access to resources of not only other co-located CDN modules but also system processes of the edge node.  On the other hand, OneThingCloud isolates different CDN modules using the container technology, and thus prevents a malicious CDN module from accessing resources belonging to other CDN modules or the edge node runtime. However, no confinement is enforced for the access to shared resources such as disk storage and bandwidth, which allows a CDN module to over-consume these resources and thus lower the quality of service of other co-located ones, e.g., extra delay in content delivery. Therefore, simply applying the cloud-based isolation solutions to edge computing may be not very feasible, especially considering the constrained resources available on edge nodes.

\subsection{Responsible Disclosures}
\label{sec:security_disclosure}
Given aforementioned security and privacy risks, We have conducted responsible disclosure to relevant parties including the OECP operators and the CDN services, for which, contact emails of these relevant parties were collected and disclosure emails have been sent. By this writing, we have received full acknowledgement from OneThingCloud and Xingyu CDN for all the reported security/privacy risks. In the replies, both mentioned that they have reproduced the reported security risks and are working to fix them. For the other parties,  we have yet to receive any concrete response.

\section{Discussion}
\label{sec:discuss}

\subject{Security recommendations.}
Below, we discuss the potential security practices to defend against aforementioned security risks. 
While defending against privacy risks, we propose EdgeTor, a protocol that features the use of edge nodes as relays to form anonymous circuits towards specific content payloads or edge nodes. Besides, we recommend that fresh threat intelligence should be integrated on edge nodes. Then, short-lived and node-specific credentials can mitigate the exposure of credentials. The details of recommendations can be found in Appendix~\ref{appendix:security_recommendations}.

\subject{The generalizability of security findings.}
As we pointed out in \S\ref{sec:risks}, some of the identified security risks widely exist in open distributed systems, and become evident when untrusted edge nodes become part of CDN services. 
For example, the low threat reputation of edge node IPs and the privacy risks against end users have also been observed in cryptojacking and residential proxies~\cite{carlin2019you, tekiner2021sok, mi2019resident, mi2021your}. Moreover, the risks related to TLS credentials are generalized to all TLS traffic in open distributed systems. Additionally, other security risks are inherent in the concept of OECPs, such as the the insufficient confinement enforced by edge platforms for their tenants. More related works can be found in Appendix~\ref{sec:related}.

\subject{The applicability of our methodology.}
We believe our methodology is applicable to OECPs beyond the two we specifically studied. Firstly, the edge node deployment framework is applicable for all OECPs that support deployment options as docker images. It follows a general pipeline that involves deploying the docker images, triggering effective edge tasks, capturing docker status and edge network traffic, and backing them up to a server. Then, the edge traffic analyzer is capable of processing raw edge traffic of any OECP. Besides, the four traffic categories we define for edge computing activities are broadly applicable to any other OECPs involved in CDN tasks. Regarding the edge security testbed, these MITM experiments are applicable to TCP edge traffic, as the tool mitmproxy is limited to handling TCP-based traffic flows. 

\section{Data Availability.}
We released the source code for all the tools developed to capture and understand edge traffic in \url{https://chasesecurity.github.io/Open_Edge_Computing_Platforms/}. Regarding the dataset in this paper, to protect privacy in the plaintext traffic and domains, it will be available  upon request and necessary background vetting.

\section{Conclusion}
\label{sec:conclusion}
As first revealed in this study, open edge computing platforms have large-scale and widely distributed edge nodes, and are being extensively adopted by CDN services in content delivery towards end users of popular online services. However, they are also inherently subject to a set of security risks, e.g., the low threat reputation of edge node IPs, the sharing of long-term credentials across edge nodes, and the lack of verifiable user consent. Most security risks can be attributed to the existence of byzantine or malicious edge nodes while the left ones root in the tension between security requirements and the complexity of enabling heterogeneous edge computing tasks on resource-constrained edge nodes. 

\section{Acknowledgments}
\label{sec:acknowledgments}
This work is jointly supported by the National Natural Science Foundation of China under Grant No. 62302473 and No. 62372268,  University of Science and Technology of China through the Innovation Fund for Young Investigators,  the Shandong Provincial Natural Science Foundation (ZR2021LZH007 and ZR2022LZH013), the Jinan City "20 New Universities" Funding Project (2021GXRC084), and the Key R\&D Program of Shandong Province, China (2024CXGC010114).

\bibliographystyle{IEEEtran}
\bibliography{ref}
\appendices

\section{Collecting and Analyzing Edge Activities}
\label{appendix:deployment}

\subject{Deployment options to choose.}
When designing this framework, the first decision we should make is about what node deployment option to choose, this is because both TipTime and OneThingCloud have offered multiple options for deploying their edge nodes, so as to accommodate for as many categories of devices as possible.
Specifically, Both of them provide edge node payload as firmware (mainly for router devices), docker images, Android apps while OneThingCloud even further supports running an edge node as a plugin of a network attached device. 
One thing to note, given observations and findings distilled from docker-based edge node deployment, it is intuitive to wonder whether they are applicable to edge nodes deployed through other options (e.g., firmware), which we have confirmed through small-scaled and short-period edge node deployment using options other than docker images.

\subject{Resource requirements.}
First, varied across OECPs, content delivery is carried out either through TCP or UDP flows. To support UDP-based content delivery, the edge node should at least be configured with an open NAT so that it can serve incoming UDP packets, e.g., via STUN-based NAT traversal. 
Furthermore, to support TCP-based content delivery, the edge node should be able to listen to a public IP address and accept incoming TCP connections. 
To satisfy both,  we configure each edge container with the \textit{host} network type, which allows it to have full access to the network of the host,  a virtual machine deployed in a public cloud with static public IP addresses attached. 

Besides, a runtime monitoring module is further equipped along the docker container of each edge node so as to capture its activities and resource usage. Particularly, given a running edge node, its resource consumption will be snapshotted every minute through the docker \textit{stats} command. 
All these runtime logs will be transmitted to a central backup server on a daily base and get cleaned up locally to free up disk space.  

\subject{Tricks for Manual Analysis.}
\label{appendix:analyze}
Below we discuss few tricks while we manually analyze the raw network traffic with Wireshark.
One is to utilize the correlation between traffic flows especially the correlation between plaintext traffic flows and encrypted ones. For instance, as revealed by our manual study, in edge nodes of TipTime, the CDN module from YunFan will be deployed as the by-default edge task. And the CDN module periodically queries the control server regarding what content to cache and where to pull the respective content payload. Once the respective content payload(e.g., an Android APK file) is cached, a log message will be sent to the logging server of the same CDN. These traffic flows are causally related. Then, since the last traffic flow for logging is in plaintext, it can thus help us to infer the semantics of the first two traffic flows which are encrypted. 

One more trick is the domain name of a traffic flow can contain important indicators regarding the underlying brand names (real-world entities). For instance, subdomains of \textit{jdcloudstatus.net} are used by remote parties to request cached content from our edge nodes through HTTPS connection. Further investigation has confirmed that this is a domain name registered under JD, Inc, one of the largest e-commerce companies in China.  
% check real-world entities from domain registration
Also, likely due to regulations in China, the owners of many domain names have registered their real-world entity information in a domain WHOIS service~\cite{beian} operated by the Ministry of Science and Technology in China. This high-fidelity domain WHOIS service was also utilized by us to confirm the real-world entity behind each traffic flow. 
% More examples will be provided shortly regarding how this domain WHOIS information has benefited our manual analysis.  
% https://beian.miit.gov.cn/#/Integrated/recordQuery

% uploading/downloading streams
Also, the difference between the downstream traffic (i.e., traffic from a remote host to our edge node) and upstream traffic for a given traffic flow can also help us distinguish traffic flows. For instance, a flow to cache content from remote servers tends to have a much higher incoming traffic volume than its outgoing traffic, while it is on the contrary for a traffic flow that pulls content from the edge node. Also, traffic flows of caching or distributing the same content payload tend to share a similar traffic volume, and can thus be correlated with high confidence.  
% For instance, a flow of downloading an Android APK file from our edge node has an outgoing traffic volume that is very similar to the incoming traffic volume of a traffic flow used to cache the same APK file.  % plaintext traffic especially some logs are very help for us to decide the activities 
Furthermore, To further validate  assumptions upon traffic flows, we also reverse engineered the programs running inside the edge node containers as well as looking into their local log files. For instance, an OneThingCloud edge node container has a log file named \textit{/tmp/wxedge.log}, which specifies what edge computing modules it has initiated, when and how the content delivery tasks start running, among others. Logs in this file has also helped us to verify the co-existence of multiple content delivery tasks that are operated by different CDN providers.

\section{Edge Nodes}
\label{appendix:edge_nodes}
\begin{figure}[h]
    \centering
    \subfigure[
        The lifetime.
        \label{subfig:cdf_lifetime}
    ]{
        \includegraphics[width=.45\columnwidth]{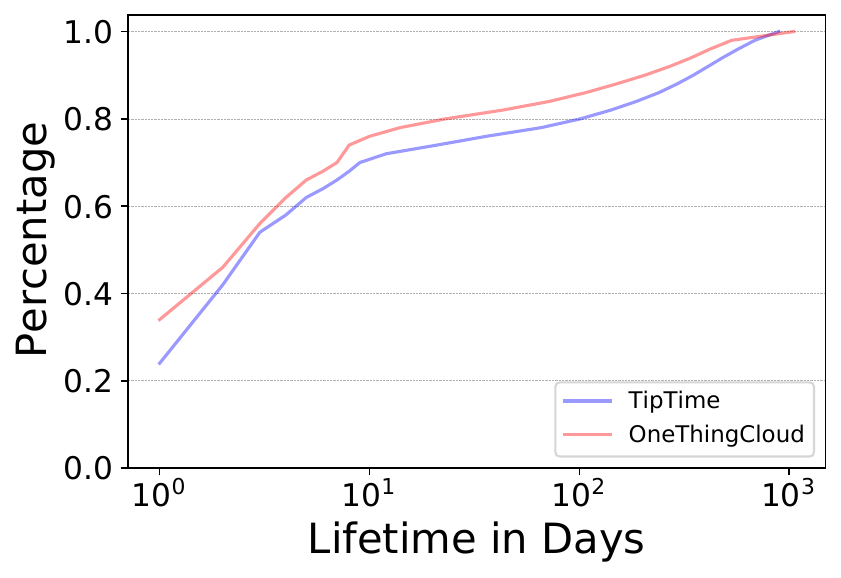}
    }
    \hfill
    \subfigure[
        The activeness time.
        \label{subfig:cdf_activetime}
        ]{
        \includegraphics[width=.45\columnwidth]{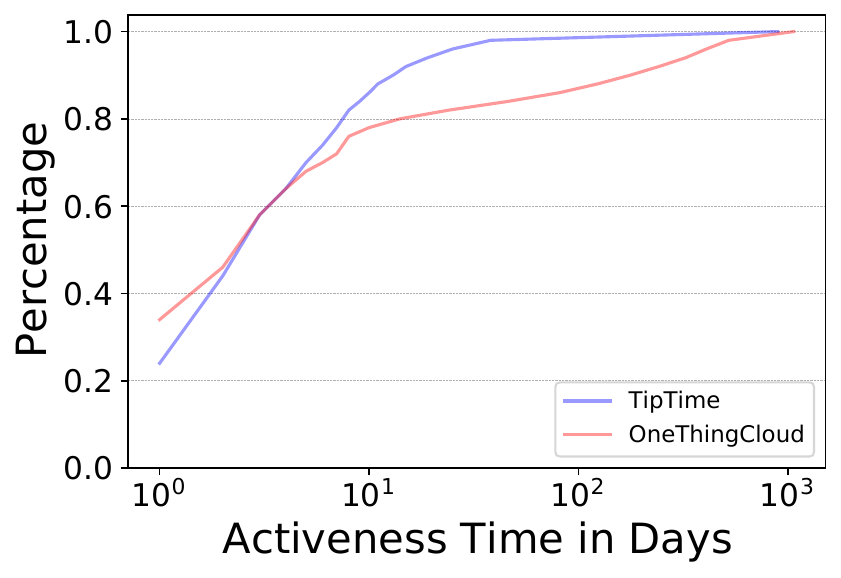}
    }
    \caption{The cumulative distribution of edge nodes over their lifetime and activeness time.}
    \label{fig:cdf_lifetime}
\end{figure}
\subject{The lifetime of edge nodes.} We also profile how long an edge node can keep alive (online) since it is first deployed. Here, we define two metrics to profile this question. The first is the lifetime $lt$ of an edge node as  $lt = date_{last} - date_{first}$, i.e., the interval in days between the first date and the last date that the FQDN of an edge node is observed in passive DNS. However, an edge node may be offline for some of the time between $date_{first}$ and $date_{last}$. We therefore define another metric, namely activeness time $at$, as $\sum_{d = first}^{last}active_d$, where $active_d$ is 1 if the edge FQDN is active on date $d$, otherwise $active_d$ is 0. 
The cumulative distribution of edge nodes across their lifetime and activeness time are presented in Figure~\ref{subfig:cdf_lifetime} and Figure~\ref{subfig:cdf_activetime} respectively. 
As we can see that both ECPs share a similar pattern that most edge nodes have a short lifetime, which is very different from traditional cloud servers. For instance, over 70\% TipTime edge nodes have a lifetime shorter than 10 days while it is 76\% for OnethingCloud edge nodes. Then, the activeness time for an edge node is even shorter. For instance, 70\% edge nodes across OECPs have an activeness time shorter than 6 days. These results echo aforementioned observations regarding the churning rate of OECPs. Two factors may jointly contribute to this non-negligible churning rate as well as the short lifetime. One is that many edge node devices are not attached to static IP addresses and thus migrate quickly across network blocks, which leads to new edge node IPs. The other factor is that many edge node devices quit the ecosystem  while new ones get recruited at a fast pace.

\section{Edge-Assisted Content Delivery}
\label{appendix:content_delivery}

\begin{table}[h]
    \centering
    \caption{The list of CDN services and content providers.}
    \footnotesize
    \begin{tabular}{
    >{\centering\arraybackslash}m{0.1\linewidth}
    >{\centering\arraybackslash}m{0.2\linewidth}
   m{0.5\linewidth}}
        \toprule
         CDN & OECP & Content Provider  \\
         \midrule
          YunFan CDN& TipTime & KuaiShou, Douyin, Baidu Cloud, PPTV, Mogen Cloud, Jingdong Cloud, Zuiyou\\
          Wangsu CDN& TipTime & Toutiao\\
          Xingyu CDN& OneThing Cloud & Zuiyou, Wasu TV, Netease, Toutiao, GiTV, imoo, Xiaomi\\
          Bilibili CDN& OneThing Cloud & Bilibili\\
          Baidu CDN& OneThing Cloud & Haokan Video, Baidu Cloud\\
          Xunlei CDN& OneThing Cloud & Xunlei \\
         \bottomrule
    \end{tabular}
    \label{tab:cdns}
\end{table}

\subject{CDN Nodes Observed in Edge Traffic.}
\label{appendix:cdn_nodes_in_edge_traffic}
Table~\ref{tab:cdn_nodes_in_traffic} presents statistical measurements for nodes of each CDN service as observed in edge traffic. Similar to Table~\ref{tab:edge_nodes_traffic} regarding edge nodes, these data points can only serve as a lower-bound estimate for the scale and distribution of CDN nodes, since not all CDN nodes would communicate with nodes under our control.
\begin{table}[h]
    \centering
    \footnotesize
    \caption{CDN nodes observed in edge traffic.}
    \label{tab:cdn_nodes_in_traffic}
    \begin{tabular}{ccccc}
        \toprule
        CDN Service & Node IPs & /8 IPv4 & ASes & Countries\\
        \midrule
        YunFan CDN & 17,585 & 51 & 46 & 1  \\
        Wangsu CDN & 1,089  & 14 & 6 &  1 \\
        Xingyu CDN & 2,818  & 32 & 5 & 1 \\
        Bilibili CDN & 43,613 & 54 & 76 & 2 \\
        Baidu CDN & 59,671 & 59  & 90 & 3 \\
        Xunlei CDN & 1,197 & 47  & 34 & 1 \\
        All & 125,187  &  61 &  120 & 4  \\
         \bottomrule
    \end{tabular}
\end{table}

\subject{End Clients of Content Delivery.}
\label{subsec:end_clients}
Table~\ref{tab:cdn_end_clients} presents the stats of end-client IPs (i.e., recipient IPs) that have received content payloads from edge nodes under our control. 
    \begin{table}
        \centering
        \footnotesize
        \caption{The stats of end-client IPs of content delivery as observed in edge traffic.}
        \label{tab:cdn_end_clients}
        \scalebox{0.83}{
        \begin{threeparttable}
            \begin{tabular}{cccccc}
                \toprule
                ECP & CDN & CP~\tnote{1} & Client IPs & /8 IPv4 & ASes  \\ 
                \midrule
                \multirow{9}{*}{TipTime} & \multirow{7}{*}{YunFan CDN} & Kuaishou  & 255,630 & 54 & 57\\
                & & Douyin & 162,028 & 63 & 80\\
                & & Baidu Cloud  & 19,114 & 53 & 50\\
                & & PPTV & 2,378 & 45 & 36\\
                & & Mogen Cloud & 1,339 & 33 & 22\\
                & & JD Cloud & 24 & 13 & 2\\
                & & Zuiyou  & 6 & 5 & 3\\
                \cline{2-6}
                & Wangsu CDN & Toutiao & 31,928 & 43 & 30\\
                & All  &  All & 465,842 & 70 & 133\\
                \hline
                \multirow{4}{*}{OneThingCloud} 
                &  Bilibili CDN & Bilibili  &  474,559 & 194 & 892\\
                & Baidu CDN & Baidu/Haokan  & 16,649 & 90 & 139\\
                & Xingyu CDN & All~\tnote{2} & 147,749 & 116 & 126\\
                & All & All & 612,220 & 211  & 935 \\  
                \hline
                All & All & All & 1,069,962 & 211 & 935 \\  
                \bottomrule
            \end{tabular}
            \begin{tablenotes}
                \item [1] CP is short for the content provider.
                \item [2] The edge-to-client traffic of Xingyu CDN uses a UDP-based proprietary protocol, for which, we are unable to infer the respective content provider.
            \end{tablenotes}        
        \end{threeparttable}}
    \end{table}

    In addition to signatures to decide client IPs and respective content platforms, some client-edge traffic flows also contain semantic fields to denote the device type and the operating system of the end-user device. For instance, end clients of Kuaishou use the \textit{x-client-info} http header to denote the device type and the operating system. For instance, the following value denotes a mobile device of Redmi Note 8 and the operating system of Android: \textit{model=Redmi Note 8 Pro;os=Android;nqe-score=74;network=WIFI;signal-strength=4;}. Leveraging these signatures, we have identified 4,367 distinct device types and 2 operating systems for 253,067 client IPs, and typical device types include iPhone14, RedmiK30Pro, V1901A (a Vivo smart phone), etc.

\section{Content Payloads Under Delivery}
\label{subsec:content_under_delivery}
Leveraging the combination of traffic signatures, plaintext logging traffic, and in-container local logs, we are allowed to not only locate content delivery flows, but also reliably decide the content types for many of these flows. 

\begin{table}[h]
    \centering
    \footnotesize
    \caption{The delivery stats for content of different types.}
    \label{tab:stats_of_content_types}
    \begin{tabular}{cccc}
        \toprule
        Content Type & Flows & Volume in GBs & Recipient IPs \\
        \midrule
        Video Streaming & 1,053,195 & 1500.40GB & 774,908\\
        Audio Streaming & 8,637 & 3.00GB & 7,683\\
        Image & 7,265 & 0.15GB & 6,966\\
        Program Files & 2,770 & 8.57GB & 2,179\\
        Compressed Files & 29,672 & 12.25GB & 18,663\\
        Others & 1,082 & 0.08GB & 1,076\\
        \bottomrule
    \end{tabular}
\end{table}

\subject{Content types.} Among content delivered in \textit{plaintext} through our edge nodes, files of 22 different types have been observed, which can be further grouped as the following content types: video streaming files (e.g., .ts/.mp4/.m4s/.m3u8 files, video/quicktime, video/mp4 types), audio streaming files (e.g., .mp3/.m4a files, audio/mpeg types), image files (e.g., .png/.webp files), program files (e.g., .nds/.json/.bspatch files, text/x-diff types), and general compressed files (e.g., .zip files). Table~\ref{tab:stats_of_content_types} lists the content types observed for content delivery traffic flows that are in plaintext. For each of these content types, we also present the volume of its content delivery traffic as well as the volume of recipient IPs. 
One thing to note, as the content types for encrypted delivery traffic flows are invisible to us,  this is only a lower-bound estimate for the diversity of content under delivery through edge nodes. 

Although many content payloads are transmitted in encrypted flows, some CDN modules will log in plaintext the HTTPS URLs for caching these content payloads. We thus manually downloaded over 19K different files by following such logged HTTPS urls, which reveal more file types like .avi files for video streaming, .flac files for audio streaming, and various program files, e.g, .apk, .pkg, .exe, .so, .rpk, .patch, etc. Furthermore, we also manually looked into many compressed files, which reveal even more file/content types, e.g., 3D model files (e.g., .frag and .vert), program files (e.g., .js, .tex, .css, .xml, .html, and .sh), serialized data files (e.g., .blob, .bundle, and .bin), machine learning models (e.g., .tflite, .model, .mlmodel, and .weights). 
Given machine learning model files observed, our further investigation reveals that they are likely used for face-relevant processing for users of Kuaishou, a major short video streaming platform in China. 
To summarize, across both OECPs, edge nodes are used to deliver content of diverse categories, which include not only traditional static web files (Javascript and media files), but also emerging content types (program files and machine learning models).

\begin{figure}
    \centering
    \includegraphics[width=.55\columnwidth]{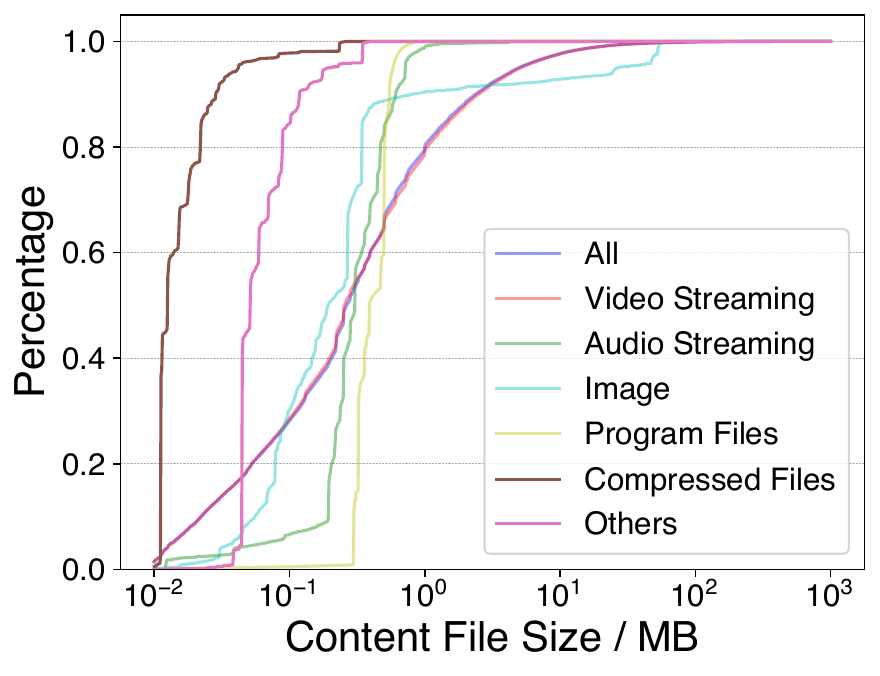}
    \caption{The cumulative distribution of content payloads over their size in MB.}
    \label{fig:cdf_content_size}
\end{figure}

\subject{Content size.} We also measured the distribution of content files over their file size. As many content delivery traffic flows are encrypted, we cannot directly know the size of such content payloads. Instead, given a content delivery flow, we use the outgoing traffic volume to approximate the size of the content under delivery, which is based on the assumption that the content file under delivery accounts for the majority of the respective outgoing traffic. Table~\ref{tab:cdf_content_size_1MB} presents the cumulative distribution of content delivery flows over the size of their outgoing traffic. And we can see that content payloads tend to be delivered in small chunks, e.g., almost 80\% content payloads under delivery have a size less than 1MB while 28\% have a size less than 100KB. Typically, the smaller the chunk size, the lower the transmission latency is.  Such a small-chunk delivery pattern is consistent with the status quo of Internet-wide content delivery especially for video streaming and live streaming~\cite{ghabashneh2020exploring}.  Then, a CDF plot is also provided in Figure~\ref{fig:cdf_content_size} to profile the full-range content size distribution.

\begin{table}
    \centering
    \footnotesize
    \caption{The cumulative distribution of content payloads over their size in MB.}
    \label{tab:cdf_content_size_1MB}
    \begin{tabular}{cccccc}
        \toprule
        Content Type & $\leq 0.05\text{MB}$ & $\leq 0.1\text{MB}$ & $\leq 0.5\text{MB}$ & $\leq 1\text{MB}$ \\
        \midrule
        All & 19.11\% & 28.15\% & 64.75\% & 79.45\% \\
        Video Streaming & 19.23\% & 28.56\% & 64.41\% & 78.52\%\\
        Audio Streaming & 4.45\% & 6.63\% & 82.37\% & 98.73\% \\
        Image & 96.21\% & 97.80\% & 100.00\% & 100.00\% \\
        Program Files & 8.84\% & 30.25\% & 88.77\% & 90.36\%\\
        Compressed Files & 0.37\% & 0.50\% & 59.80\% & 99.99\% \\
        Others & 45.47\% & 84.47\% & 100.00\% & 100.00\% \\
        \bottomrule
    \end{tabular}
\end{table}

\subject{Recipient distribution.} As discussed above, our edge nodes have delivered content to 1,129,132 distinct recipient IPs. We further profiled the distribution of these recipient IPs over the volume of content that each of them has received, in an attempt to further understand how our edge nodes are used in content delivery. Here, we measure the volume of content delivery through the outgoing traffic volume of these flows. As shown in Table~\ref{tab:cdf_recipient_ips_size_1MB}, the content payloads received by most recipient IPs are of a small volume, e.g., 83.85\% recipient IPs have received no more than 1MB content from our edge nodes. 
Such an observation suggests that a typical end-user in edge-assisted content delivery may concurrently request small content chunks of a large file from many CDN nodes.  
We also surfaced out top recipient IPs with the largest volume of content delivered, as listed in Table~\ref{tab:top_recipient_ips}. In contrast with the overall pattern, each of  the top 10 recipients  has over 1GB content delivered. Our further investigation shows  that they are all CDN nodes of Bilibili rather than end users.

\begin{table}
    \centering
    \footnotesize
    \caption{The cumulative distribution of recipient IPs over the size of content delivery smaller than 1MB.}
    \label{tab:cdf_recipient_ips_size_1MB}
    \begin{tabular}{cccccc}
        \toprule
        Edge Platform & $\leq 0.05\text{MB}$ & $\leq 0.1\text{MB}$ & $\leq 0.5\text{MB}$ & $\leq 1\text{MB}$ \\
        \midrule
        Tiptime & 6.12\% & 8.89\% & 55.03\% & 79.38\% \\
        OneThing Cloud & 36.22\% & 48.62\% & 77.55\% & 86.90\% \\
        Both & 23.77\% & 32.17\% & 68.28\% & 83.85\% \\
        \bottomrule
    \end{tabular}
\end{table}

\begin{table}
    \centering
    \footnotesize
    \caption{Top 10 recipient IPs with the largest volume of content received from our edge nodes.}
    \label{tab:top_recipient_ips}
    \begin{tabular}{cccc}
        \toprule
        IP & Edge Platform & Volume in MB & Traffic Flows\\
        \midrule
        1.93.47.215 & OneThing Cloud & 3.36 GB & 8 \\
        120.229.86.167 & OneThing Cloud & 2.66 GB & 9 \\
        112.0.216.70 & OneThing Cloud & 1.90 GB & 1 \\
        183.211.129.248 & OneThing Cloud & 1.67 GB & 2 \\
        182.99.7.26 & OneThing Cloud & 1.63 GB & 9 \\
        112.111.104.62 & OneThing Cloud & 1.58 GB & 3 \\
        180.212.71.12 & OneThing Cloud & 1.52 GB & 8 \\
        114.253.242.246 & OneThing Cloud & 1.51 GB & 5 \\
        111.79.194.94 & OneThing Cloud & 1.50 GB & 3 \\
        58.22.211.92 & OneThing Cloud & 1.50 GB & 2 \\
        \bottomrule
    \end{tabular}
\end{table}

\section{Extra Data for Threat Reputation of Edge Node IPs}
\label{appendix:threat_reputation}

\begin{table}[h]
    \centering
    \footnotesize
    \caption{The threat reputation of edge node IPs when measured against malicious traffic flows (MTFs) of the same date. 
    }
    \label{tab:stats_private_threat_same_date}
    \scalebox{0.87}{
    \begin{threeparttable}[online]
        
    \begin{tabular}{llcccc}
        \toprule
        Edge  & Group  & w MTFs & $\ge 5$ MTFs & $\ge 10$ MTFs & Median MTFs\\
        \midrule
        \multirow{2}{*}{TipTime} 
        & $\text{Edge}_{\text{Traffic}}$  &  6.15\% &     3.30\% &      2.52\% &            6  \\
         & $\text{CDN}_{\text{Traffic}}$ &   6.03\% &     3.26\% &      2.52\% &            6   \\
        \multirow{2}{*}{OneThing} 
        & $\text{Edge}_{\text{Traffic}}$ &  2.94\% &     1.77\% &      1.32\% &            6  \\ 
         & $\text{CDN}_{\text{Traffic}}$ & 4.07\% &     2.25\% &      1.71\% &            6 \\ 
        \hline
         \multirow{2}{*}{Both}
           & $\text{Edge}_{\text{Traffic}}$ & 5.13\% &     2.82\% &      2.14\% &            6 \\ 
         & $\text{CDN}_{\text{Traffic}}$ & 4.69\% &     2.57\% &      1.97\% &            6  \\ 
        \bottomrule
    \end{tabular}
   \end{threeparttable}}
\end{table}
\subject{MTFs reports from the proprietary threat intelligence platform. }
Table~\ref{tab:stats_private_threat_same_date} presents the ratio of edge node IPs that have one or more MTFs observed. One thing to note,  an MTF is counted for an edge node IP only if it is captured on a date when the IP is serving as an edge node.  

\begin{table}
    \centering
    \footnotesize
    \caption{Statistics of edge nodes' malicious activities as captured by VirusTotal: \textit{Mal} denotes malicious.}
    \label{tab:stat_threat_vt}
    \begin{tabular}{clccc}
        \toprule
      Edge  & Group  & Mal & W Mal URLs \ & W Malware \\
        \midrule
        \multirow{3}{*}{TipTime} 
        & $\text{Edge}_{\text{Traffic}}$ &3.99\% &      3.96\% &     0.81\%\\
        & $\text{CDN}_{\text{Traffic}}$ &3.62\% &      3.59\% &     0.73\% \\
        & $\text{Edge}_{\text{DNS}}$ &2.75\% &      2.73\% &     0.91\%\\
        \hline
        \multirow{3}{*}{OneThing} 
        & $\text{Edge}_{\text{Traffic}}$ &2.33\% &      2.33\% &     0.22\% \\
        & $\text{CDN}_{\text{Traffic}}$&3.01\% &      2.99\% &     0.46\%\\
         & $\text{Edge}_{\text{DNS}}$&3.07\% &      3.04\% &     1.31\%\\
        \hline
         \multirow{3}{*}{Both} 
          & $\text{Edge}_{\text{Traffic}}$& 3.47\% &      3.44\% &     0.62\%\\
         & $\text{CDN}_{\text{Traffic}}$ &3.21\% &      3.19\% &     0.55\%\\
         & $\text{Edge}_{\text{DNS}}$ & 2.78\% &      2.76\% &     0.94\%\\
        \bottomrule
    \end{tabular}
\end{table}
\subject{Threat reports from VirusTotal.} 
We also queried Virustotal~\cite{vt_api} with the aforementioned groups of edge node IPs. The VirusTotal reports for a given IP typically contain the malicious URLs the IP has ever hosted, as well as the set of malware the IP has associated with. And there are three types of associations between an IP and a malware. One is embedding wherein the IP is embedded in the payload for a given malware. The second is communicating which denotes that a malware has ever communicated with the given IP. The third association type is hosting which means the IP is found to have ever hosted the given malware for distribution.  Among these three association categories, embedding and communicating are weak indicators to judge maliciousness of an IP address. Instead, hosting malware is commonly recognized as a strong maliciousness indicator. Therefore, we exclude the weak indicators when analyzing VirusTotal reports, and an IP is considered as malicious only when it hosts either malicious URLs or malware. 

As shown in Table~\ref{tab:stat_threat_vt}, a non-negligible fraction of edge node IPs are considered as malicious by VirusTotal for either hosting malicious URLs or distributing malware payloads. Particularly, among edge node IPs directly observed in edge traffic ($\text{Edge}_{\text{Traffic}}$), 3.47\% are detected as malicious by VirusTotal. We then look into top IPs that are most reported by VirusTotal, and we observe that 49 out of the top 50 IPs used to be bots of the Mozi IoT botnet since they have one or more URLs detected for Mozi payload distribution.  For instance, \textit{183.15.89.96}, a TipTime edge node observed in traffic, used to distribute Mozi payloads and has 20 URLs detected by VirusTotal, e.g., \textit{http://183.15.89.96:38713/Mozi.m}, and \textit{https://183.15.89.96:38713/mozi.a}.
Among IPs having malicious URLs detected, over 25\% have such kinds of Mozi payload URLs.

\section{Security Recommendations.} 
\label{appendix:security_recommendations}

\subsubject{Defending against privacy risks.} While the large-scaled untrusted edge nodes incur non-negligible risks, they also create an opportunity to address aforementioned privacy risks. Taking lessons from anonymity networks such as Tor, we propose \textit{EdgeTor}, a protocol that features the use of edge nodes as relays to form anonymous circuits towards specific content payloads or edge nodes. In the Tor network, a circuit towards a traffic destination is set up by the end user in advance by separately contacting three or more Tor relays. In contrast, in EdgeTor, a circuit can be set up by either the CDN control server or the edge control server, which thus avoids any modifications to the end-user devices. 

Next, we illustrate this protocol using the case of edge-assisted content delivery. Specifically, a CDN control server should dynamically populate each on-edge CDN module with a routing table which specifies next hop to forward an incoming request of a content payload identified by its URL. Also, the number of intermediate hops in a circuit is configurable and CDN services can dynamically tune this value so as to achieve the best trade-offs between privacy and efficiency. Also, one should expect extra latency as more hops will be traversed for each content delivery traffic flow. However, as the edge nodes are large-scaled and widely distributed, clusters can be formed wherein each pair of edge nodes are close to each other with a very low latency. Then, a circuit can be formed with relays all from the same low-latency cluster. Through this protocol, as long as edge nodes are not heavily colluded, the IP of the end user will not be visible to the edge node serving the requested content while the edge node directly connected to the end user has no knowledge of what content is transmitted. Based on the guidelines above, we will implement the EdgeTor framework in the future works.

\subsubject{The integration of threat intelligence.} As revealed above, edge node IPs can be considerably involved in malicious activities, and fresh threat intelligence should be integrated into the operation of edge nodes and content delivery modules, so as to timely exclude suspicious edge nodes from serving edge computing tasks, e.g., delivering  content towards end users. 

\subsubject{Storing only short-lived and node-specific credentials on edge nodes.} Also, credentials distributed to edge nodes should be short-lived, periodically updated, and more importantly, differ across edge nodes. Also, as the trusted execution environment (TEE) is increasingly supported by more computing devices, the TEE support, if available on edge nodes, can be used to generate and store the credentials discussed above. All these measures can effectively mitigate the exposure of valid credentials to attackers. 

\section{Related Works}
\label{sec:related}

\subject{Measurements on closed edge computing platforms.} Xu, et al.~\cite{xu2021cloud} conducted the first empirical measurement on a public and operational edge platform, and have revealed multiple interesting observations regarding performance and resource management. For instance, the median round-trip time (RTT) between an end device and the nearest edge node is 10.5ms, which is 1.89× (19.8ms) faster than the nearest cloud site. Also, the studied edge platform was found to have been used in various content delivery activities. 
 However, our study differs from \cite{xu2021cloud} in multiple aspects. First of all, the research subjects are different with regards to who owns the edge computing node. In the platform studied in \cite{xu2021cloud}, edge nodes are exclusively deployed and operated by the platform operator, while edge nodes of the two OECPs under our study are contributed by third parties. We thus name the platform studied in \cite{xu2021cloud} as \textit{closed} edge computing platform since all edge nodes are centrally deployed  by the platform itself. Secondly, \cite{xu2021cloud} focused more on the quality of service and workload allocation while our study cares more about the security and privacy implications. 

\subject{Security studies on edge computing.} Previous security studies on edge computing ~\cite{park2019streambox, mo2020darknetz, islam2023confidential} focus on proposing and experimenting security solutions for various edge computing scenarios. Particularly, Park et al.~\cite{park2019streambox} explored the adoption of trusted execution environment (TEE) for privacy-preserving video analytics in edge devices, while Mo et al.~\cite{mo2020darknetz} studied how to protect deep neural network models running in an edge device from potential membership inference attacks. To the best of our knowledge, our work is the first empirical security study on two open and operational edge computing platforms. 
\end{document}